\documentclass{article}
\usepackage[english]{babel}

\usepackage[margin=1in]{geometry}

\usepackage[utf8]{inputenc} 
\usepackage[T1]{fontenc}    
\usepackage{hyperref}       
\usepackage{url}            
\usepackage{booktabs}       
\usepackage{amsfonts}       
\usepackage{nicefrac}       
\usepackage{microtype}      
\usepackage{lipsum}		
\usepackage{graphicx}
\usepackage[numbers]{natbib}
\usepackage{doi}
\usepackage{amsmath}
\usepackage{mathtools}
\usepackage{tabularx,ragged2e,booktabs,caption}
\usepackage{amsthm}

\usepackage{amssymb}
\usepackage{amsmath}
\usepackage{commath}
\usepackage{tikz}
\usetikzlibrary{positioning}
\newcommand{\ind}{\perp\!\!\!\!\perp}

\newcommand{\mycomment}[1]{}
\usepackage{hyperref}
\usepackage[parfill]{parskip}
\usepackage{multirow}

\usepackage{cleveref}
\crefname{remark}{rmk}{rmks}
\usepackage{subcaption}

\usepackage{indentfirst}
\usepackage[margin=1in]{geometry}

\usepackage[utf8]{inputenc} 
\usepackage[T1]{fontenc}    
\usepackage{hyperref}       
\usepackage{url}            
\usepackage{booktabs}       
\usepackage{amsfonts}       
\usepackage{nicefrac}       
\usepackage{microtype}      
\usepackage{lipsum}		
\usepackage{graphicx}
\usepackage[numbers]{natbib}
\usepackage{doi}
\usepackage{mathtools}
\usepackage{amsmath}
\usepackage{tabularx,ragged2e,booktabs,caption}
\usepackage{tikz}
\usetikzlibrary{positioning}
\usepackage[makeroom]{cancel}
\usetikzlibrary{shapes,arrows}
\usepackage{caption}
\usetikzlibrary{matrix,shapes,arrows,positioning,chains}
\newtheorem{prop}{Proposition}[section]

\tikzset{
decision/.style={
    ellipse,
    draw,
    text width=10em,
    text badly centered,
    inner sep=3pt
},
block/.style={
    rectangle,
    draw,
    text width=14em,
    text centered,
    rounded corners
},
cloud/.style={
    draw,
    ellipse,
    minimum height=2em
},
descr/.style={
    fill=white,
    inner sep=2.5pt
},
connector/.style={
    -latex,
    font=\scriptsize
},
rectangle connector/.style={
    connector,
    to path={(\tikztostart) -- ++(#1,0pt) \tikztonodes |- (\tikztotarget) },
    pos=0.5
},
rectangle connector/.default=-2cm,
straight connector/.style={
    connector,
    to path=--(\tikztotarget) \tikztonodes
}
}

\makeatletter
\newcommand*{\centernot}{%
  \mathpalette\@centernot
}
\def\@centernot#1#2{%
  \mathrel{%
    \rlap{%
      \settowidth\dimen@{$\m@th#1{#2}$}%
      \kern.5\dimen@
      \settowidth\dimen@{$\m@th#1=$}%
      \kern-.5\dimen@
      $\m@th#1\not$%
    }%
    {#2}%
  }%
}
\makeatother

\newcommand{\independent}{\perp\mkern-9.5mu\perp}
\newcommand{\notindependent}{\centernot{\independent}}

\date{} 					
\begin{document}
\title{Choosing the Right Approach at the Right Time: A Comparative Analysis of Causal Effect Estimation using Confounder Adjustment and Instrumental Variables}

\author{{Roy S. Zawadzki}\thanks{Corresponding Author: zawadzkr@uci.edu}\\
	Department of Statistics\\
	University of California, Irvine\\
	Irvine, CA\\
	\and
	{Daniel L. Gillen}\\
	Department of Statistics\\
	University of California, Irvine\\
	Irvine, CA\\}


\maketitle

\begin{abstract}
\noindent In observational studies, potential unobserved confounding is a major barrier in isolating the average causal effect (ACE). In these scenarios, two main approaches are often used: confounder adjustment for causality (CAC) and instrumental variable analysis for causation (IVAC). Nevertheless, both are subject to untestable assumptions and, therefore, it may be unclear which assumption violation scenarios one method is superior in terms of mitigating inconsistency for the ACE. Although general guidelines exist, direct theoretical comparisons of the trade-offs between CAC and the IVAC assumptions are limited. Using ordinary least squares (OLS) for CAC and two-stage least squares (2SLS) for IVAC, we analytically compare the relative inconsistency for the ACE of each approach under a variety of assumption violation scenarios and discuss rules of thumb for practice. Additionally, a sensitivity framework is proposed to guide analysts in determining which approach may result in less inconsistency for estimating the ACE with a given dataset. We demonstrate our findings both through simulation and by revisiting Card's analysis of the effect of educational attainment on earnings, which has been the subject of previous discussion on instrument validity. The implications of our findings on causal inference practice are discussed, providing guidance for analysts to judge whether CAC or IVAC may be more appropriate for a given situation.
\end{abstract}

\section{Introduction}
A common goal in observational studies is to estimate the average causal effect (ACE) of a treatment or exposure on a specific outcome such as the effect educational attainment on earnings. Due to the exposure not being randomized, the presence of confounders may bias estimates of the ACE. Confounders are factors that influence both the level of exposure (or treatment assignment) and the outcome. If uncontrolled for, confounders can create extraneous differences between the exposure groups that make it difficult to isolate the causal effect. The effectiveness of confounder adjustment for causality (CAC), however, is contingent on the untestable assumption that all possible confounders, or suitable proxies, are present in the dataset and have been accounted for appropriately. In other words, we cannot use observed data to prove that the CAC is consistent for the ACE. A simple example is if household income, a known confounder for educational attainment and earnings, and any proxies were missing from the dataset we would like to analyze. Without \textit{a priori} knowledge that household income was a confounder, we would not be able to ascertain that the CAC assumptions were violated.

An alternative approach that may avoid concerns about unobserved confounders is the instrumental variable (IV) approach, or instrumental variable analysis for causation (IVAC). An IV is defined by three main conditions: (i) it influences the treatment assignment (relevance), (ii) it is not a cause of the outcome after conditioning on treatment assignment (exclusion restriction), and (iii) it is not associated with unobserved confounders (independence). In the event that we have a variable that satisfies these conditions, we could then use variation in the IV as a proxy for variation in the treatment and measure the effect on the outcome. Importantly, by (iii), the variation in the IV is independent from unobserved confounding and, therefore, unlike CAC, IVAC does not require appropriately accounting for all possible confounding to consistently estimate the ACE.\citep{baiocchi_tutorial_2014}

When choosing to use CAC over the IVAC, and vice-versa, we trade one set of untestable assumptions for another. For CAC, we cannot prove that we have accounted for all confounders while, for IVAC, we cannot prove we have a valid IV. For example, proving the exclusion restriction requires us to establish the lack of a direct relationship between the IV and outcome, or a null result, which is not possible with data. Furthermore, to be consistent for the ACE, the IVAC must meet certain untestable conditions surrounding treatment effect heterogeneity. We therefore have no guarantee that under either approach our produced estimate is consistent for the ACE. Despite this, we remain interested in estimating the ACE and subsequently direct our efforts towards attempting to estimate a parameter with the least possible distance from the ACE. In this vein, the key question and central premise of the current manuscript focuses on addressing the following question in practice: under the potential violation of the untestable assumptions, when is the parameter estimated by CAC closer to the ACE than that of IVAC, and vice versa?

In the literature, there exist general guidelines surrounding whether CAC is more appropriate than IVAC.\citep{baiocchi_tutorial_2014,zawadzki_frameworks_2023} Generally, we contend that analysts should weigh whether the potential degree of unobserved confounding outweighs the potential for violations in the IV assumptions. There is, however, little by way of theoretical research that directly compares the two approaches to assess these trade-offs. To examine these trade-offs between CAC and IVAC, we focus on the use of ordinary least squares (OLS) and two-stage least squares (2SLS), respectively, to estimate the causal effect in each paradigm for two main reasons. First, OLS and 2SLS are the most commonly used methodologies for each approach and, therefore, our findings would be readily applicable to a large group of analyses. Second, linear functional forms will allow us to give intuitive and tractable closed-form results for relative inconsistencies. Then, we will provide a sensitivity framework to guide analysts in determining whether the inconsistency of 2SLS is more than that of OLS, and vice versa.

Alternative estimators for CAC and IVAC include using non-linear machine learning (ML) models such as double machine learning or targeted minimum loss estimation approach.\citep{chernozhukov_doubledebiased_2018,van_der_laan_targeted_2010} Though flexible modeling may protect against functional form misspecification, they are far from immune to inconsistency due to the assumption violations we will study. Yet, non-linear approaches will carry additional complexity that will make it difficult to analytically quantify the impact of potential violations in closed form. Thus, for our purposes, we will utilize OLS and 2SLS to estimate causal effects under a linear data generating mechanism with the general intuitions gleaned translating to non-linear settings.

In order to more succinctly express the relative performance of OLS and 2SLS under assumption violations, we focus on the scenario where, for a given variable, we must decide on whether the variable should be adjusted for as a confounder in OLS, used as an IV in 2SLS, or not incorporated into the analysis at all. Note that we allow confounders to be adjusted for in 2SLS -- an IV is only used in the first stage whereas confounders would need to be present in both the first and second stage.

To our knowledge, there is no existing theoretical literature directly comparing the trade-offs of pursuing CAC versus IVAC, though authors have considered the impact of assumption violations in both settings. The first area of literature relevant to the themes of our paper is bias amplification, which refers to the fact that using certain variables as confounders may increase pre-existing bias due to unobserved confounding. In the linear setting, an IV is a bias amplifier.\citep{bhattacharya_instrumental_2007,wooldridge_should_2016,pearl_class_2012,stokes_causal_2022} In these papers, the authors compare the consistency for the ACE with and without adjusting for an IV. Pearl (2012) provides an extension where there is an imperfect IV (in that the exclusion restriction is violated) and shows that under certain conditions, adjusting for the imperfect IV may actually reduce bias. Similar to our paper, he presents his results with both linear structural equations and directed acyclic graphs (DAG) edge-weights to aid understanding of the the trade-offs. Nevertheless, he does not address whether this variable may be more appropriately used in 2SLS.

The bias amplification literature and, by extension, our findings have important implications on applied practice. In particular, the rise of data-driven variable selection approaches for the propensity score, or probability of receiving the intervention, in confounder methods and first-stage for IV analysis. Though IVs, by definition, influence the treatment assignment, they may amplify bias if included in the model for the propensity score.\citep{bhattacharya_instrumental_2007} For IV methods, though a variable may not be a perfect IV it may still be worth using it as such. In addition, data-driven modeling of the first-stage may, for example, shrink to zero an important confounder used to achieve the IV assumptions. The gain in the strength of the first-stage may, however, offset the penalty incurred by omitted variable bias in 2SLS. As it stands, these intuitions are difficult to incorporate into variable selection procedures. With this work, we seek to elucidate these complicated scenarios.

In another area of the literature, there has been some work related to comparing OLS to 2SLS under the violation of IV assumptions. It is a well-known result that if an IV is poorly predictive of the intervention (i.e. weak), then small violations in the exclusion restriction and independence assumption can lead to large inconsistencies for the IV estimand.\citep{bound_problems_1995,wooldridge_econometric_2010}. In addition, to assess the independence assumption, one may compare the impact of intentionally omitting an observed confounder on OLS and 2SLS in order to compare the sensitivity of OLS to that of 2SLS in estimating the ACE.\citep{brookhart_preference-based_2007} The main assumption behind this procedure is that the impact of omitting an observed confounder on the consistency of OLS and 2SLS is similar to that of omitting a correlated unobserved confounder. A similar assumption and "benchmarking" procedure will be used in our sensitivity analysis. 

There are several procedures in the literature regarding sensitivity analyses for violations in IV assumptions. For example, Cinelli and Hazlett (2022) provide a compelling framework and visualization scheme for omitted variable bias in 2SLS based on several partial $R^2$ measures.\citep{cinelli_omitted_2022} Their framework addresses the question of how large the impact of an unobserved confounder would have to be in order to qualitatively change the inferential conclusions of a study, which covers both violations of the exclusion restriction and independence assumptions. A similar procedure to this is the E-value.\citep{vanderweele_sensitivity_2017} While we use many of the same tools -- notably benchmarking unobserved $R^2$ measures with observed data from Cinelli and Hazlett (2022) -- we do not focus on this sensitivity analysis paradigm of hypothesis testing but instead consider the relative inconsistencies of CAC and IVAC for the ACE.

Another complication to IVAC lies in treatment effect heterogeneity. In this setting, Imbens and Angrist (1994) state that, under monotonicity conditions, IVAC identifies the local average causal effect (LACE) or the causal effect of the "compliers" subpopulation (those whose treatment assignment varies with the IV).\citep{imbens_identification_1994} If the factors that determine compliance also cause treatment effect heterogeneity, then the LACE may not be equal to the ACE. Hartwig et al. (2020) and Wang and Tchetgen Tchetgen (2018) give clear explanations of the assumptions needed for the LACE to equal the ACE.\citep{hartwig_average_2022,wang_bounded_2018} Essentially, the heterogeneity between the treatment and outcome should be independent of both the IV and the effect modification between the treatment and outcome. For the ease of parameterization in our paper, we will use Wang and Tchetgen Tchetgen's notion that one of two conditions must be met: (i) no unmeasured confounders are additive effect modifiers of the relationship of both the instrument and treatment or (ii) no unmeasured confounders are additive effect modifiers the treatment and the outcome.

The rest of the paper is organized as follows. First, we  provide the general model setting of interest and introduce relative notation. We then present results regarding the consistency of no adjustment, OLS, and 2SLS for the scenarios of an exclusion restriction violation, independence violation, and treatment effect heterogeneity relevant to IV estimation with and without covariates. In all scenarios, we consider unobserved confounding and isolate the impact of individual assumption violations (e.g. both an exclusion restriction and independence violation). Following this, we present a sensitivity analysis procedure based on partial $R^2$ and benchmarking unobserved quantities with observed quantities. The goal of this procedure is to give the analyst relevant information to assess the plausibility of whether it may be more appropriate to adjust for a variable in OLS or 2SLS. Next, in simulations, we verify our closed-form results and demonstrate the use of the sensitivity analysis procedure in a variety of scenarios. Then, we apply the procedure to the analysis the effect of educational attainment on earnings conducted in Card (1993), which has been the subject of subsequent analyses over the validity of the IV utilized.\citep{card_using_1993,cinelli_omitted_2022} We conclude with a discussion regarding the implications of our closed form results on the practice of causal inference and, additionally, provide further guidance on how to use our sensitivity analysis procedure.

\section{Notation and Set-Up}

Our goal is to estimate the ACE of some continuous treatment $X$ on an outcome $Y$, denoted as $\beta_1 = \frac{\partial}{\partial x}E[Y|do(x)]$ in Pearl's notation.\citep{pearl_causality_2009} We depict the causal relationships in Figure \ref{fig:dag_simple}, a DAG with edge weights $c_0,\dots,c_3$. We further consider the following structural equations where $E[\epsilon_1|U,Z] = 0$ and $E[\epsilon_2|X,U] = 0$:
\begin{equation}\label{eq:simple_first}
    X = \alpha_1U + \alpha_2Z + \epsilon_1,
\end{equation}
\begin{equation}\label{eq:simple_second}
    Y = \beta_1X + \beta_2U + \epsilon_2.
\end{equation}

\begin{figure}[h!]
    \centering
    \begin{tikzpicture}[thick]
        \node (1) at (0,0) {Z};
        \node (2) at (2,0) {X};
        \node (3) at (5,0) {Y};
        \node (5) at (3.5,1) {U};

        \path [->] (1) edge node[above]{$c_3$} (2);
        \path [->] (2) edge node[above]{$c_0$} (3);
        \path [->] (5) edge node[above]{$c_1$} (2);
        \path [->] (5) edge node[above]{$c_2$} (3);
    \end{tikzpicture}
    \caption{A directed acyclic graph with one confounder and one instrumental variable.}
    \label{fig:dag_simple}
\end{figure}

Given that all variables in Figure (\ref{fig:dag_simple}) are standardized to have mean $0$ and variance $1$, the edge weights are equivalent to the coefficients in Eqs. (\ref{eq:simple_first}) and (\ref{eq:simple_second}). For example, the ACE, $\beta_1$, is the same as the edge weight $c_0$. In addition, $\alpha_1 = c_1$, $\alpha_2 = c_3$, $\beta_1 = c_0$, and $\beta_2 = c_2$. This equivalence will be helpful in visually expressing assumptions surrounding different scenarios. Furthermore, the edge weights are correlations and are bounded between $-1$ and $1$.

Throughout, we assume the stable unit treatment value assumption (SUTVA) of consistency and no interference. Because $U$ is unobserved, we must estimate the the following reduced form regression where the subscript $R$ indicates these are the values related to the reduced regression
\begin{equation}
    Y = \beta_1^{R}X + \epsilon_2^{R}.
\end{equation}

$U$ is a confounder because it both influences $X$ and $Y$, and, therefore, because we have failed to block the path through $U$, $\hat{\beta}_1^{R}$ is inconsistent for the ACE. Alternatively, to estimate the ACE, we may utilize $Z$, which is a valid IV if
\begin{enumerate}
    \item $\alpha_2 > 0$ (relevance)
    \item $Z$ is not a cause of $Y$ conditional on $X$ (exclusion restriction)
    \item $Z$ is not influenced by any unaccounted for confounders such that $Z \notindependent Y | X$ (independence).
\end{enumerate}

Supposing that there is no treatment effect heterogeneity or that the conditions of Hartwig et al. (2020) or Wang and Tchetgen Tchetgen (2018) \citep{hartwig_average_2022,wang_bounded_2018} are met, we have $\beta_{IV} = \frac{Cov(Z,Y)}{Cov(Z,X)} = \frac{\beta_1\alpha_1}{\alpha_1} = \beta_1$, and hence 2SLS will provide a consistent estimate of the ACE. The first equality comes from the definition of the LACE under a continuous treatment and outcome and, under the aforementioned conditions, the LACE is equal to the ACE.

We are interested in estimating and comparing the following estimands: the causal effect i.e. Eq. (\ref{eq:causal_estimand}), one that omits $Z$ i.e. Eq. (\ref{eq:no_adj_estimand}), one that uses $Z$ as a confounder i.e. Eq. (\ref{eq:adj_estimand}), and one that utilizes $Z$ as an IV i.e. Eq.(\ref{eq:2sls_estimand}):
\begin{equation}\label{eq:causal_estimand}
    A_1 = \frac{\partial}{\partial x}E[Y|do(x)] = c_0,
\end{equation}
\begin{equation}\label{eq:no_adj_estimand}
    A_2 = \frac{\partial}{\partial x}E[Y|x],
\end{equation}
\begin{equation}\label{eq:adj_estimand}
    A_3 = \frac{\partial}{\partial x}E[Y|x,z],
\end{equation}
\begin{equation}\label{eq:2sls_estimand}
    A_4 = \frac{Cov(Y,Z)}{Cov(X,Z)} = \frac{\frac{\partial}{\partial z}E[Y|z]}{\frac{\partial}{\partial z}E[X|z]}.
\end{equation}


We define the degree of inconsistency for $A_1$ of estimates of $A_2$, $A_3$, and $A_4$ as a set of absolute differences: $\lambda_2 = |A_1 - A_2|$, $\lambda_3 = |A_1 - A_3|$, and $\lambda_4 = |A_1 - A_4|$. Our goal is to compare the magnitude of $lambda_2$, $\lambda_3$, and $\lambda_4$. One approach "performs better" than the other if the respective $\lambda$ is smaller. For example, 2SLS performs better than OLS if $\lambda_4 > \lambda_3$. 

As a simple example of the types of calculations and comparisons that we will do in the next section, we can use the conditions of Figure \ref{fig:dag_simple}. From Pearl (2012), we have that $A_2 = c_0 + c_1c_2$ by Wright's rules of path analysis and $A_3 = c_0 + c_2\frac{\partial}{\partial x}E[U|x,z] =  c_0 + \frac{c_1c_2}{1-c_3^2}$.\citep{pearl_class_2012} As a result of adjusting for $Z$ as a confounder, we have bias amplification that increases with the strength of the IV (i.e. the magnitude of $c_3$). In addition, we have that $\hat{A}_4 \overset{p}{\to} c_0$. The proof of this is in the first section of the Appendix. Further, by Chebyshev's inequality and assuming finite variances hold, we also have that $\hat{A}_2 \overset{p}{\to} c_0 + c_1c_2$ and $\hat{A}_3 \overset{p}{\to} c_0 + \frac{c_1c_2}{1-c_3^2}$. Thus, $\lambda_3 = \frac{c_1c_2}{1-c_3^2} > \lambda_2 =  c_1c_2 > \lambda_4 = 0$ or, in words, 2SLS performs better than no adjustment, which performs better than OLS.

\section{Trade-offs Under Violations in Instrumental Variable Assumptions}

In this section, we present results for the trade-offs between confounder and IV methods for three scenarios: (i) violation of the exclusion restriction assumption, (ii) violation of the independence assumption, and (iii) treatment effect heterogeneity. In all scenarios, $U$ is unobserved, which provides the realistic setting where we may be motivated to use 2SLS due to the concern of unobserved confounding. For ease of exposition, we first derive the quantities of interest without adjustment for observed confounding. We then present the quantities with these observed covariates. For ease of comparison of the quantities of interest, we further assume all regression slope parameters are positive throughout this section and we will handle the general case in the sensitivity analysis portion. Unless otherwise stated, all proofs for the propositions in this section can be found in the Appendix.

\subsection{Exclusion Restriction Violation}

Figure \ref{fig:excl_rest_dag} directly reproduces Figure 2 from Pearl (2012) and presents a violation of the exclusion restriction assumption for $Z$ if $c_{ER} \ne 0$. We use this quantity to denote the degree of violation. By traditional logic, one would define $Z$ as a confounder because it is a cause both of $X$ and $Y$ and use it as such. It is not, however, unequivocally true that one should use $Z$ as a confounder. To see this, suppose we have the following structural equations: 
\begin{equation}\label{eq:er_eq1}
    X = c_1U + c_3Z + \epsilon_3
\end{equation}
\begin{equation}\label{eq:er_eq2}
    Y = c_0X + c_2U + c_{ER}Z + \epsilon_4.
\end{equation}

\begin{figure}[h!]
    \centering
    \begin{tikzpicture}[thick]
        \node (1) at (0,0) {Z};
        \node (2) at (2,0) {X};
        \node (3) at (5,0) {Y};
        \node (5) at (3.5,1) {U};

        \path [->] (1) edge node[above]{$c_3$} (2);
        \path [->] (2) edge node[above]{$c_0$} (3);
        \path [->] (5) edge node[above]{$c_1$} (2);
        \path [->] (5) edge node[above]{$c_2$} (3);
        \path [->] (1) edge[bend right=30] node[above] {$c_{ER}$} (3);
    \end{tikzpicture}
    \caption{Exclusion Restriction Violation using $Z$ as an IV.}
    \label{fig:excl_rest_dag}
\end{figure}

\begin{prop}\label{rmk:er}
Under the conditions of Eqs. (\ref{eq:er_eq1}) and (\ref{eq:er_eq2}), $\hat{A}_2 \overset{p}{\to} c_0+c_1c_2+c_3c_{ER}$, $\hat{A}_3 \overset{p}{\to} c_0 + \frac{c_1c_2}{1-c_3^2}$, and $\hat{A}_4 \overset{p}{\to} c_0 + \frac{c_{ER}}{c_3}$.
\end{prop}

The convergence results for $A_2$ and $A_3$ can be found in Pearl (2012) so we omit them. The proof for $A_4$ can be found in the Appendix. We see that adjusting for $Z$ decreases inconsistency compared to not adjusting for $Z$ if $\frac{c_{ER}}{c_3} \ge \frac{c_1c_2}{1-c_3^2}$. This inequality could be difficult to attain if the instrument is strong.\citep{pearl_class_2012} Interestingly, the left term of this inequality is the inconsistency of 2SLS and thus the IV being strong is relatively advantageous for the use of $Z$ as an IV in 2SLS. We can re-arrange the inequality between $A_3$ and $A_4$ as $\frac{c_{ER}(1-c_3^2)}{c_3} \ge c_1c_2$ where $c_1c_2$ indicates the impact of unobserved confounding in the relationship between $X$ and $Y$. Here, it becomes more clear that the strength of the IV can be large enough such that the degree of exclusion restriction violation (i.e. $c_{ER}$) is offset and is smaller than the impact of unobserved confounding.

\begin{figure}[h]
    \centering
    \caption{Analytical comparison of no adjustment, OLS adjusting for the IV, and 2SLS using the IV under varying IV strength and degree of exclusion restriction violation.}\par\medskip
    \includegraphics[scale=0.5]{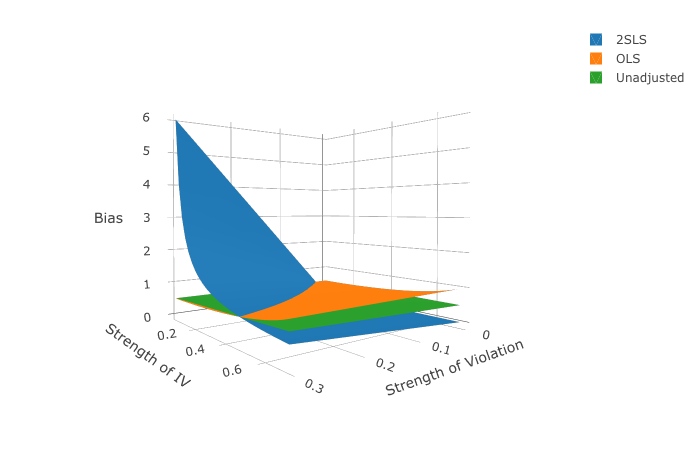}
    \label{fig:er_viz}
\end{figure}

The trade-offs between $A_2$, $A_3$, and $A_4$ can be visualized with a 3-D contour plot. In Figure \ref{fig:er_viz}, letting $c_0 = 0.3, c_1 = 0.7,$ and $c_2 = 0.7$, we can vary the values of $c_3$ and $c_{ER}$. Note the plausible coefficient values for $c_3$ and $c_{ER}$ are restricted due to the requirement that the variances for the variables to sum to one (see "Notes about Simulations" section in the Appendix). This image gives us the visual intuition that when the IV is stronger, moderate violations in the exclusion restriction violation do not preclude the use of $Z$ as an IV. Furthermore, adjusting for $Z$ will be inferior compared to not adjusting for $Z$. When the IV is weak, 2SLS predictably performs poorly in all cases.

\subsection{Independence Violation}

\begin{figure}[h]
    \centering
        \begin{tikzpicture}[thick]
        \node (1) at (0,0) {Z};
        \node (2) at (2,0) {X};
        \node (3) at (5,0) {Y};
        \node (5) at (2.5,2) {U};

        \path [->] (1) edge node[above]{$c_3$} (2);
        \path [->] (2) edge node[above]{$c_0$} (3);
        \path [->] (5) edge node[right]{$c_1$} (2);
        \path [->] (5) edge node[above]{$c_2$} (3);
        \path [->] (5) edge node[above]{$c_I$} (1);
    \end{tikzpicture}
    \caption{$Z$ is Correlated with an Unobserved Confounder $U$}
    \label{fig:iv_conf_dag}
\end{figure}

Figure \ref{fig:iv_conf_dag} represents one violation of the independence assumption where the residual confounder $U$ is a cause of $Z$. Here, $c_{I}$ represents the degree of violation or, in this case, the effect of $U$ on $Z$. Alternative specifications create a new confounder that is only associated with $Z$ but not $X$. Nonetheless, our parameterization provides a useful case where both the independence assumption is violated and there is confounding in the relationship between $Z$ and $X$. For structural equations, we can re-use Eqs \ref{eq:simple_first} and \ref{eq:simple_second} as well as add an extra structural equation given by
\begin{equation}\label{eq:z_u_violation}
    Z = c_IU + \epsilon_5.
\end{equation}

\begin{prop}\label{rmk:uc}
Under Eqs. (\ref{eq:simple_first}), (\ref{eq:simple_second}), and (\ref{eq:z_u_violation}), $\hat{A}_2 \overset{p}{\to} c_0 + c_1c_2+c_2c_3c_I$, $\hat{A}_3 \overset{p}{\to} c_0 + \frac{c_1c_2(1-c_I^2)}{1-(c_3+c_1c_I)^2}$, and $\hat{A}_4 \overset{p}{\to} c_0 + \frac{c_2c_I}{c_3 + c_1c_I}$.
\end{prop}

Establishing the convergence result for $A_2$ is a straightforward application of Wright's path analysis so only the proofs for $A_3$ and $A_4$ are provided in the Appendix.\citep{wright_method_1934} To better interpret these quantities, we can think about both paths on the DAG and remaining variance after orthogonalizing variables via the Frisch-Waugh-Lovell Theorem (FWL).\citep{lovell_simple_2008} Looking at the pathways, $c_1c_2$ is the backdoor path from $X$ to $Y$ through $U$, $c_2c_3c_I$ is the backdoor path through $Z$, and $c_2c_I$ is the path from $Z$ to $Y$ via $U$. $c_3 + c_1c_I$ is the unconditional correlation between $Z$ and $X$, which includes both the direct and backdoor path via $U$. $1-c_I^2$ depicts the remaining (stochastic) variation in $Z$ after orthogonalizing $U$ while $1-(c_3 + c_1c_I)^2$ is the remaining variance in $X$ after orthogonalizing $Z$.

\begin{figure}[h!]
    \centering
    \caption{Analytical comparison of no adjustment, OLS adjusting for the IV, and 2SLS using the IV under varying IV strength and degree of the independence violation.}\par
    \includegraphics[scale=0.5]{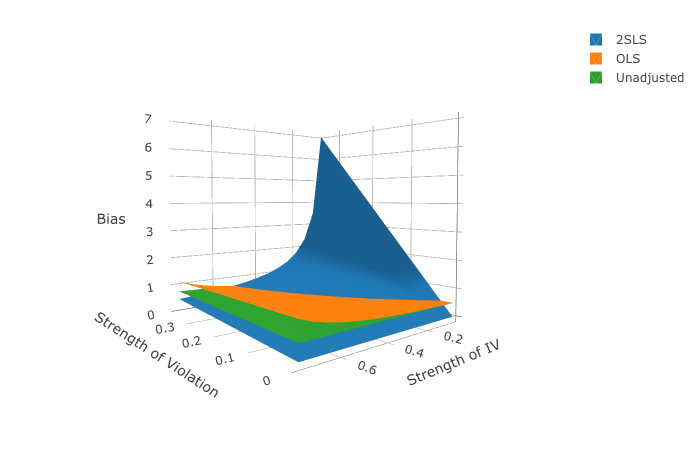}
    \label{fig:uc_viz}
\end{figure}

Of particular interest is the trade-off between using $Z$ in 2SLS and using $Z$ in OLS. We find that adjusting for $Z$ is superior to using $Z$ for 2SLS if $\frac{c_I}{c_3 + c_1c_I} > \frac{c_1(1-c_I^2)}{1-(c_3+c_1c_I)^2}$. We can begin to interpret this inequality with the reoccurring theme that if the IV is strong then attaining this inequality is more difficult: a strong IV will cause $c_3+c_1c_I$ to be large which inflates the inconsistency in $A_3$ while it decreases the consistency for $A_4$. The remaining variation in $Z$ not caused by $U$, or $c_I$, is an important quantity because as $c_I$ decreases, the inconsistency in $A_3$ will increase while the inconsistency in $A_4$ will decrease. In this sense, a simple sensitivity analysis procedure could be to benchmark the variation in the IV that is explained by the covariates. One could use this benchmark to conjecture how much of the variation in the IV is explained by an unobserved confounder. If this quantity is small then one could plausibly assume a fair amount of variation in $Z$ free from unobserved confounding and, thus, $c_I$ is small. These trade-offs can be visualized in Figure \ref{fig:uc_viz} where $c_0 = 0.3, c_1 = 0.7,$ and $c_2 = 0.7$.

\subsection{Treatment Effect Heterogeneity}

We return the setting of Figure \ref{fig:dag_simple} where we have a perfect IV but now we omit the edge weights and introduce treatment effect heterogeneity. In this case, we do not know of any literature that gives us a salient way to to represent treatment heterogeneity using DAGs and edge weights. Therefore, the nuance is in the structural equations:
\begin{equation}\label{eq:first_stage_em}
    X = \alpha_1Z + \alpha_2U + \alpha_3ZU + \epsilon_1,
\end{equation}
\begin{equation}\label{eq:second_stage_em}
    Y = \beta_1X + \beta_2U + \beta_3 XU + \epsilon_2.
\end{equation}

The estimand of interest remains the ACE or the average of the individual treatment effects, which is denoted by $\beta_1$. This parameterization is consistent with a violation of Wang and Tchetgen Tchetgen (2018) assumptions 5a and 5b if $\alpha_3 \ne 0$ and $\beta_3 \ne 0$. The key parameter for measuring the degree of "assumption violation" is $\alpha_3$ because we aim to quantify how large an unobserved confounder needs to be in order to modify the effect on $Z$ in the first-stage to render $\lambda_4 > \lambda_3$ and $\lambda_4 > \lambda_2$, or that IVAC is inferior to CAC.

We note that in this scenario, $U$ is extended to be a composite variable that includes unobserved confounders but, additionally, unobserved effect modifiers. That is, a variable that only affects the outcome and thus contribute to the magnitude of $\beta_2$ and $\beta_3$ but not $\alpha_2$ and $\alpha_3$, and vice versa. This allows us to be more flexible in that we do not require all unobserved variables to be both confounders and effect modifiers but perhaps only effect modifiers.

\begin{prop}\label{rmk:em}
Under the conditions of Eqs. (\ref{eq:first_stage_em}) and (\ref{eq:second_stage_em}) as well as assuming $E[U^3] = 0$, $\hat{A}_{2} \overset{p}{\to} \beta_1 + \alpha_2\beta_2 + 2\alpha_1\alpha_3\beta_3, \hat{A}_{3} \overset{p}{\to} \beta_1 + \frac{\alpha_2\beta_2 + \alpha_1\alpha_3\beta_3}{1-\alpha_1^2},$ and $\hat{A}_{4} \overset{p}{\to} \beta_1 + \frac{\alpha_3\beta_3}{\alpha_1}$
\end{prop}

When comparing the relative trade-off between not adjusting for $Z$ and adjusting for $Z$, the latter is inferior if $\frac{\alpha_2\beta_2 + \alpha_1\alpha_3\beta_3}{1-a_1^2} > \alpha_2\beta_2 + 2\alpha_1\alpha_3\beta_3$. Unlike the case of no effect modification, it is not always true that adjusting for an IV will amplify bias. Specifically, by rearranging terms we see the inequality holds if $2\alpha_1^2 + \frac{\alpha_2\beta_2}{\alpha_3\beta_3} > 1$. If $|\alpha_1| \gtrapprox 0.7$, or the IV is strongly associated with $X$, then this inequality will hold; however, this would not be the case if the IV is sufficiently weak and the multiplication of the coefficients for the interactions are larger than the multiplication of the coefficients representing the main effect.

\begin{figure}[h!]
    \centering
    \caption{Analytical comparison of no adjustment, OLS adjusting for the IV, and 2SLS using the IV under varying IV strength and degree of the treatment effect heterogeneity assumption.}\par\medskip
    \includegraphics[scale=0.5]{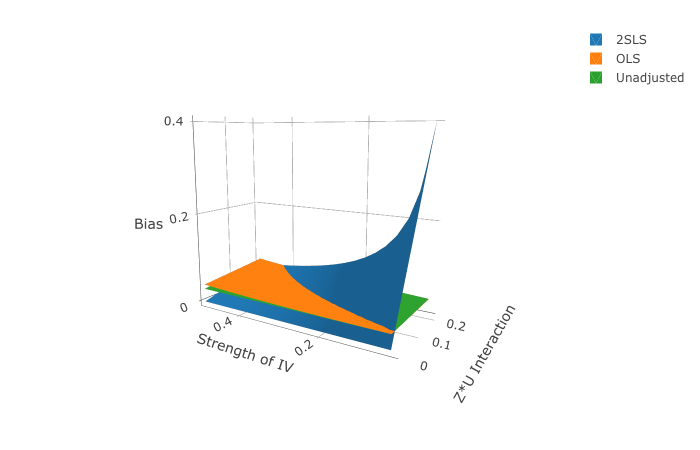}
    \label{fig:em_viz}
\end{figure}

In the more realistic case, if an analyst is aware $Z$ is an IV and assuming that the IV was sufficiently strong, the main point of concern surrounds whether the LACE estimate is more inconsistent than the ACE estimate from the unadjusted OLS.  This notion is true if $\frac{\alpha_3\beta_3}{\alpha_1} > \alpha_2\beta_2 + 2\alpha_1\alpha_3\beta_3$ or $\frac{\alpha_1\alpha_2\beta_2}{\alpha_3\beta_3} + 2\alpha_1^2 < 1$. Firstly, a strong IV, $|\alpha_1| \gtrapprox 0.7$, precludes this inequality from being attained. Alternatively, this inequality could fail to be attained if the ratio of main effects to interaction effects, scaled by the IV strength, is large. Setting $\alpha_2 = 0.15$, $\beta_1 = 0.1$, $\beta_2 = 0.2$, and $\beta_3 = 0.1$, we can visualize the above inequalities in Figure \ref{fig:em_viz}.

A final takeaway from these results is that the strength of an IV has implications on the the OLS inconsistencies even though it is independent from $U$ (e.g. via the term $2\alpha_1\alpha_3\beta_3$ in $A2$). Therefore, accounting for any present IVs, even if never utilized, is important for understanding the degree of inconsistency present in confounder methodology. The results of the previous three Propositions are summarized in Table \ref{tab:summary_aim1}.

\begin{table}[!htbp]
    \centering
    \caption{Summary of Results}
    \begin{tabular}{p{2.5in} |c|c|c}
        \toprule
        \hline
        \bf Scenario & $A_2$ & $A_3$ & $A_4$ \\
        \midrule
        \hline
        No Violation\\
        \begin{tikzpicture}[thick]
        \node (1) at (0,0) {Z};
        \node (2) at (2,0) {X};
        \node (3) at (5,0) {Y};
        \node (5) at (3.5,1) {U};

        \path [->] (1) edge node[above]{$c_3$} (2);
        \path [->] (2) edge node[above]{$c_0$} (3);
        \path [->] (5) edge node[above]{$c_1$} (2);
        \path [->] (5) edge node[above]{$c_2$} (3);
    \end{tikzpicture}\\
    \vbox{\begin{equation*}
    X = c_1 U + c_3 Z + \epsilon_1
    \end{equation*}
    \begin{equation*}
    Y = c_0 X + c_2 U + \epsilon_2
    \end{equation*}}
    & $c_0+c_1c_2$ & $c_0 + \frac{c_1c_2}{1-c_3^2}$ & $c_0$\\
    \hline
    Exclusion Restriction Violation\\
    \begin{tikzpicture}[thick]
        \node (1) at (0,0) {Z};
        \node (2) at (2,0) {X};
        \node (3) at (5,0) {Y};
        \node (5) at (3.5,1) {U};

        \path [->] (1) edge node[above]{$c_3$} (2);
        \path [->] (2) edge node[above]{$c_0$} (3);
        \path [->] (5) edge node[above]{$c_1$} (2);
        \path [->] (5) edge node[above]{$c_2$} (3);
        \path [->] (1) edge[bend right=30] node[above] {$c_{ER}$} (3);
    \end{tikzpicture}\\
    \vbox{\begin{equation*}
    X = c_1U + c_3Z + \epsilon_3
    \end{equation*}
    \begin{equation*}
    Y = c_0X + c_2U + c_{ER}Z + \epsilon_4.
    \end{equation*}}
    & $c_0+c_1c_2+c_3c_{ER}$ & $c_0 + \frac{c_1c_2}{1-c_3^2}$ & $c_0 + \frac{c_{ER}}{c_3}$\\
    \hline
    Independence Assumption \\
    \begin{tikzpicture}[thick]
        \node (1) at (0,0) {Z};
        \node (2) at (2,0) {X};
        \node (3) at (5,0) {Y};
        \node (5) at (2.5,2) {U};

        \path [->] (1) edge node[above]{$c_3$} (2);
        \path [->] (2) edge node[above]{$c_0$} (3);
        \path [->] (5) edge node[right]{$c_1$} (2);
        \path [->] (5) edge node[above]{$c_2$} (3);
        \path [->] (5) edge node[above]{$c_I$} (1);
    \end{tikzpicture}\\
    \vbox{\begin{equation*}
    X = c1_1U + c_3Z + \epsilon_1
    \end{equation*}
    \begin{equation*}
    Y = c_0X + c_2U + \epsilon_2
    \end{equation*}
    \begin{equation*}
    Z = c_IU + \epsilon_5.
    \end{equation*}}
    \vbox{}
    & $c_0 + c_1c_2+c_2c_3c_I$ & $c_0 + \frac{c_1c_2(1-c_I^2)}{1-(c_3+c_1c_I)^2}$ & $c_0 + \frac{c_2c_I}{c_3 + c_1c_I}$\\
    \hline
    Treatment Effect Heterogeneity\\
    \vbox{\begin{equation*}
        X = \alpha_1Z + \alpha_2U + \alpha_3ZU + \epsilon_1,
    \end{equation*}
    \begin{equation*}
        Y = \beta_1X + \beta_2U + \beta_3 XU + \epsilon_2.
    \end{equation*}}
    & $\beta_1 + \alpha_2\beta_2 + 2\alpha_1\alpha_3\beta_3$ & $\beta_1 + \frac{\alpha_2\beta_2 + \alpha_1\alpha_3\beta_3}{1-a_1^2}$ & $\beta_1 + \frac{\alpha_3\beta_3}{\alpha_1}$\\
    \bottomrule
    \end{tabular}
    \label{tab:summary_aim1}
\end{table}

\subsection{Adding Additional Confounders}

In the vast majority of data analyses, an analyst will have access to observed confounders that will be adjusted for to mitigate confounding both in OLS and 2SLS. Temporarily, we consider a single observed confounder, $W$, that is continuous and has mean $0$ and variance $1$. If we have multiple confounders, $\{W_1,W_2,...W_J\}$, that are in the same form as $W$, we can simply redefine $W$ as some function of the confounders, $f(W_1,W_2,...W_J)$. For example, this might take a form linear combination derived from the first principal component of the $W$ matrix. Therefore, the edge weights and regression coefficients will be the joint effect of the confounders and our updated results can accommodate multiple covariates. Depending on the which equation $W$ is in, it can additionally represent the propensity score (first stage) or prognostic score (second stage) that can be estimated using ML models under a partial linear model that relaxes assumptions the functional form of covariates.\citep{robinson_root-n-consistent_1988,chernozhukov_doubledebiased_2018} In this work, for simplicity, we will not discuss such ways of modeling $W$.

Because we would like to use $W$ to benchmark relationships involving $U$, we must update the DAGs and structural equations such that the $W$ has a similar relationship to $U$ in each assumption violation. As a consequence, our previous results will change. In all cases, $W$ is orthogonal to $U$ or, in other words, $U$ represents confounding in ACE unrelated to $U$. With the exception of $A_1$, the quantities of interest are updated to

\begin{align}
    A_2 &= \frac{\partial}{\partial x}E[Y|x,w]\\
    A_3 &= \frac{\partial}{\partial x}E[Y|x,z,w]\\
    A_4 &= \frac{Cov(Y,Z|W)}{Cov(X,Z|W)} = \frac{\frac{\partial}{\partial z}E[Y|z,w]}{\frac{\partial}{\partial z}E[X|z,w]}.
\end{align}

\subsubsection{Exclusion Restriction}

Figure \ref{fig:larger_DAG_er} serves as an example of a covariate that has no direct impact on $Z$ or $U$. Nevertheless, if we were to condition upon $X$ and nothing else, $U$ would no longer be independent of $W$ or $Z$ due to the collider effect. A collider effect induces an association between two variables that point to (i.e. cause) a single variable that has been conditioned.\citep{pearl_causality_2009} However, conditioning upon $W$ breaks this association.

\begin{figure}[h!]
    \centering
        \begin{tikzpicture}[thick]
        \node (1) at (0,0) {Z};
        \node (2) at (2,0) {X};
        \node (3) at (5,0) {Y};
        \node (4) at (3.5,-1) {W};
        \node (5) at (3.5,1) {U};

        \path [->] (1) edge node[above]{$c_3$} (2);
        \path [->] (2) edge node[above]{$c_0$} (3);
        \path [->] (5) edge node[above]{$c_1$} (2);
        \path [->] (5) edge node[above]{$c_2$} (3);
        \path [->] (1) edge[bend right=70] node[above] {$c_{ER}$} node[above]{$c_{ER}$} (3);
        \path [->] (4) edge node[below]{$c_5$} (2);
        \path [->] (4) edge node[below]{$c_6$} (3);
    \end{tikzpicture}
    \caption{A DAG with observed confounder and violation of exclusion restriction ($W$).}
    \label{fig:larger_DAG_er}
\end{figure}

The updated structural equations are
\begin{equation}\label{eq:er_eq1_larger}
    X = c_1U + c_3Z + c_5W + \epsilon_3
\end{equation}
\begin{equation}\label{eq:er_eq2_larger}
    Y = c_0X + c_2U + c_{ER}Z + c_6W + \epsilon_4.
\end{equation}

\begin{prop}\label{rmk:er_larger}
    Under Eqs. (\ref{eq:er_eq1_larger}) and (\ref{eq:er_eq2_larger})$, \hat{A}_2 \overset{p}{\to} c_0+\frac{c_1c_2+c_3c_{ER}}{1-c_5^2}$ and $\hat{A}_3 \overset{p}{\to} c_0 + \frac{c_1c_2}{1-c_3^2-c_5^2}$
\end{prop}

The proof is very similar to that of Proposition \ref{rmk:er} so it is omitted. As expected, adjusting for $W$ leads to decreased variance in $X$, which leads to a higher proportional contribution of $U$, at the benefit of eliminating the backdoor path via $c_5c_6$. For 2SLS, the results for $A_4$ are not affected because the assumption violations vis-a-vis using $Z$ as an IV are not influenced by $W$. Nevertheless, in practical settings, adjusting for $W$ will usually increase precision of $\hat{A}_4$. 

\subsubsection{Independence}

\begin{figure}[h!]
    \centering
    \begin{tikzpicture}[thick]
        \node (1) at (0,0) {Z};
        \node (2) at (2,0) {X};
        \node (3) at (5,0) {Y};
        \node (5) at (2,2) {U};
        \node (6) at (2,-2) {W};
        
        \path [->] (1) edge node[above]{$c_3$} (2);
        \path [->] (2) edge node[above]{$c_0$} (3);
        \path [->] (5) edge node[right]{$c_1$} (2);
        \path [->] (5) edge node[above]{$c_2$} (3);
        \path [->] (5) edge node[above]{$c_{I}$} (1);
        \path [->] (6) edge node[right]{$c_5$} (2);
        \path [->] (6) edge node[above]{$c_6$} (3);
        \path [->] (6) edge node[above]{$c_7$} (1);
    \end{tikzpicture}
    \caption{$W$ mimics $U$ in the DAG.}
    \label{fig:iv_conf_dag_larger}
\end{figure}

Besides introducing $W$ as a confounder in the $X-Y$ relationship, Figure \ref{fig:iv_conf_dag_larger} extends $W$ to be a confounder in the $Z-Y$ and $Z-X$ relationships. Therefore, the validity of $Z$ as an IV is contingent on conditioning upon both $W$ and $Z$. Thus, the structural equations are updated to

\begin{equation}\label{eq:z_u_violation_larger}
    Z = c_{I}U + c_7W + \epsilon_5
\end{equation}
\begin{equation}\label{eq:simple_first_larger}
    X = c_1U + c_3Z + c_5W + \epsilon_1,
\end{equation}
\begin{equation}\label{eq:simple_second_larger}
    Y = c_0X + c_2U + c_6W + \epsilon_2.
\end{equation}

The results for all quantities, $A_2$, $A_3$, and $A_4$, can now be updated per the following proposition with the proof detailed in the Appendix:

\begin{prop}\label{rmk:uc_larger}
    Under Eqs. (\ref{eq:simple_first_larger}), (\ref{eq:simple_second_larger}), and (\ref{eq:z_u_violation_larger}):
    \begin{align}
        \hat{A}_2 \overset{p}{\to} c_0+\frac{c_1c_2+c_2c_3c_I}{1-(c_5+c_3c_7)^2},
    \end{align}
    \begin{align}
        \hat{A}_3 \overset{p}{\to} c_0 + \frac{\frac{c_1c_2(1-c_7^2-c_I^2)}{1-c_7^2}}{1-(c_5+c_3c_7)^2-(1-c_7^2)(c_3+\frac{c_1c_I}{1-c_7^2})^2},
    \end{align}
    \begin{align}
        \hat{A}_4 \overset{p}{\to} c_0 + \frac{c_2c_I}{(1-c_7^2)(c_3+\frac{c_1c_I}{(1-c_7^2)})}.
    \end{align}
\end{prop}

These results bear some resemblance to Proposition \ref{rmk:uc} when we did not have $W$ present. For $A_2$, in the numerator, because $W$ does not mitigate the influence of $U$ in the DAG, we still have two backdoor paths from $X$ to $Y$ that go through $U$. Meanwhile, in the denominator the variance of $X$ is reduced via controlling for $W$, which has a direct path to $X$ as well as an indirect path via $Z$. 

For $A_3$, the quantity is similar conceptually to our findings in Proposition \ref{rmk:uc} except that we must additionally account for controlling for $W$. In the numerator, $c_1c_2(1-c_7^2-c_I^2)$ represents the magnitude of unobserved confounding reduced (i.e. multiplied) by the exogenous variance of $Z$ due to there no longer being a backdoor path to $Y$ via $Z$. We must account for the cost of adjusting $W$ therefore this quantity is amplified (i.e. divided) by $1-c_7^2$, the variance in $Z$ free of $W$. In the denominator, we have the remaining variance of $X$ after adjusting for $W$ and $Z$. The first subtracting term represents the unconditional $R^2$ between $X$ and $Z$ while the second takes the variation of $Z$ free of $W$ and multiplies it by the partial $R^2$ between $Z$ and $X$, adjusting for $W$. Because we haven't adjusted for $U$, the backdoor path via $U$ remains and, furthermore, because $W$ does not mitigate this, $W$ essentially acts like a bias amplifier for $c_1c_I$. Lastly, $A_4$ represents the association between $Z$ and $X$ via $U$ as well as the variation exogenous from $W$ directly from $Z$.

\subsubsection{Treatment Effect Heterogeneity}

Because we require our observed confounder to be of the same form of $U$ for benchmarking purposes, we will set $W$ as both an effect modifier of the treatment on the outcome and of the instrument on the treatment assignment with the following structural equations:
\begin{equation}\label{eq:first_stage_em_extended}
    X = \alpha_1Z + \alpha_2U + \alpha_3ZU + \alpha_4 W + \alpha_5 ZW + \epsilon_1,
\end{equation}
\begin{equation}\label{eq:second_stage_em_extended}
    Y = \beta_1X + \beta_2U + \beta_3 XU + \beta_4 W + \beta_5 XW + \epsilon_2.
\end{equation}

The presence of $XW$, which is observed but still endogenous, means that in OLS, we must adjust for it and in 2SLS, must provide an an additional IV. In particular, we will choose $ZW$. Therefore we modify the quantities of interest to
\begin{align}
    A_2 &= \frac{\partial}{\partial x}E[Y|x,w,xw]\\
    A_3 &= \frac{\partial}{\partial x}E[Y|x,z,w,xw]\\
    A_4 &= \frac{\partial}{\partial \hat{x}}E[Y|\hat{x},\widehat{xw},w]
\end{align}
where $\hat{x}$ and $\widehat{xw}$ represent the fitted values from using $Z$, $ZW$, and $W$ as regressors for $X$ and $XW$, respectively. Note that now the main effect is obtained after orthogonalizing $XW$ and $W$, or $\widehat{XW}$ and $W$ in 2SLS, which we can find via FWL by treating $XW$ as any other covariate. Therefore, the corresponding interpretation of the main effect is when $W=0$, or we are at the average value of the covariate due to centering the variable.

\begin{prop}\label{rmk:em_larger}
Under Eqs. (\ref{eq:first_stage_em_extended}) and (\ref{eq:second_stage_em_extended}) as well as $E[U^3] = 0$, $E[W^3] = 0$, $E[W^4] = 3$, and $E[U^4] = 3$ we have $\hat{A}_2 \overset{p}{\to} \beta_1 + \frac{\alpha_2\beta_2 + 2\alpha_1\alpha_3\beta_3}{1-\alpha_4^2-\frac{4\alpha_1^2\alpha_5^2}{1 + \alpha_4^2 + 2\alpha_5^2}}$ and $\hat{A}_4 \overset{p}{\to} \beta_1 + \frac{\alpha_1\alpha_3\beta_3 - \frac{2\alpha_1\alpha_3\alpha_5^2\beta_3}{\alpha_1^2 + \alpha_5^2}}{\alpha_1^2 + \alpha_5^2 - \frac{4\alpha_1^2\alpha_5^2}{(\alpha_1^2 + \alpha_5^2})}$.
\end{prop}

The proof for Proposition \ref{rmk:em_larger} is shown in the appendix. The interpretation of $A_2$ is consistent with Proposition \ref{rmk:em} with the denominator reflecting the fact that we are adjusting for $W$ and $XW$, which reduces the remaining variance of $X$. For $A_4$, because we are no longer computing the ratio of coefficients the interpretation is not directly comparable to Proposition \ref{rmk:em}. Nevertheless, we can see the influence of the $Z-U$ interaction on the inconsistency and observe that because $XU$ is correlated with $XW$, adjusting for $XW$ will reduce the influence of $XU$, hence the subtraction terms.

One may notice that we do not consider the convergence of $A_3$ and this is because we only wish to compare $A_2$ and $A_4$. The reason for this stems from our discussion of Proposition \ref{rmk:em} where, assuming $sign(\alpha_2\beta_2) = sign(\alpha_3\beta_3)$ for sake of simplicity, bias amplification will hold in the effect modification case if $2\alpha_1^2 + \frac{\alpha_2\beta_2}{\alpha_3\beta_3} > 1$. When we have a strong IV, or $|\alpha_1| \gtrapprox 0.7$, then this inequality holds. If we instead lower the strength of the IV to where we must consider $\frac{\alpha_2\beta_2}{\alpha_3\beta_3} > 1 - 2\alpha_1^2$, we would require the multiplication interaction effects to be at least as large as the main effects with this requirement increasing as the IV strength increases. We argue that in this circumstance one should avoid using $Z$ altogether because concern over this ratio would imply that the IV is weak and the LACE is likely to be far from the ACE due to large heterogeneity. Thus, a comparison between OLS adjusting for $Z$ and 2SLS is not warranted.

\section{Sensitivity Analysis}

In this section, we use the closed form derivations of the previous section to develop a set of sensitivity analysis procedures that provide analysts with information surrounding whether OLS or 2SLS may be more appropriate given the observed data and hypothesized assumption violations. We focus on comparing $A_3$ and $A_4$, or using $Z$ as a confounder versus as an IV, by graphically presenting the relative inconsistency $\phi = \frac{\lambda_4}{\lambda_3}$ as measured by the degree of IV assumption violations and unobserved confounding. 

The graphical depiction of the relative inconsistencies is largely motivated by Cincelli and Hazlett (2020) \citep{cinelli_making_2020} and Cincelli and Hazlett (2022)\citep{cinelli_omitted_2022} where they use a set of partial $R^2$'s to characterize how large the assumption violations in confounder and 2SLS analyses must be to render statistically significant results null. Instead of focusing on hypothesis testing, however, we examine how large unobserved confounding and IV assumption violations could be in order for move us away from ambivalence over the choice of methodology (i.e. $\phi = 1$) either towards 2SLS ($\phi > 1$) or OLS ($\phi < 1$). Quantifying this $\phi$ one IV assumption at a time, we build a detailed picture of the strengths and limitations relating to an analysis. Following this, like Cincelli and Hazlett, we use benchmarking to estimate the unobserved quantities contained within the closed-form derivations across a variety of scenarios. In the remainder of this section, we will focus on the sensitivity analysis procedure in the case of the exclusion restriction. Then, using the same principles, present the results for the independence and heterogeneity assumption violations more briefly. 

\subsection{Exclusion Restriction Violation}

For the exclusion restriction, we have that 
\begin{align}
    \phi = \bigg|\frac{c_{ER}}{c_3}\bigg|\bigg|\frac{c_1c_2}{1-c_3^2-c_5^2}\bigg|^{-1}.
\end{align}
Both $c_3$ and $1-c_3^2-c_5^2$ are observed quantities and can be directly estimated via OLS where $c_3 = \frac{\partial}{\partial z}E[X|z,w]$ and $1-c_3^2-c_5^2 = 1-R^2_{X \sim W + Z}$, or the residual variation in the model. $c_{ER}$ measures the degree of the exclusion restriction violation and, crucially, we cannot directly identify it with the reduced model $\beta^R_{ER} = \frac{\partial}{\partial z}E[Y|x,w,z]$ because conditioning on $X$ induces a collider effect between $Z$ and $U$ thus $\beta^R_{ER} = c_{ER} - \frac{c_1c_2c_3}{1-c_3^2-c_5^2}$ (see appendix). Therefore, we can substitute for $c_{ER}$ resulting in $\phi = \bigg|\frac{\beta^R_{ER} + \frac{c_1c_2c_3}{1-c_3^2-c_5^2}}{c_3}\bigg|\bigg|\frac{c_1c_2}{1-c_3^2-c_5^2}\bigg|^{-1}$. The quantification of $c_{ER}$ requires knowing $sign(c_1c_2)$, which is unobserved. As a result, we will calculate $\phi$ separately for the case when $sign(c_1c_2) = 1$ and when when $sign(c_1c_2) = -1$ 

We now turn to $c_1$ and $c_2$, which together quantify the degree of unmeasured confounding. In order to reasonably benchmark these relationships, we must make the following assumption: for the set of observed confounders with cardinality $J$ that make up the composite confounder $W$ in our notation, the magnitude of the $W_j-X$ relationship and $W_j-Y$ relationship for $j \in \{1,2,...J\}$ are similar to the magnitude of the $U-X$ and $U-Y$ relationship, respectively. To be conservative, we take the largest such magnitude where we the benchmark $c_1$ via $c_1^B = \max_j \frac{\partial}{\partial w_j}E[X|z,w_1,w_2,...,w_J]$ and similar for $c_2$. Of course, the degree of unobserved confounding could be different than the observed benchmark. To mitigate this, we may additionally add a multiplier, for example $M \times c_1^B$, if we believe the observed confounders underestimate ($M < 1$) or overestimate ($M > 1$) the degree of unobserved confounding. For instance, if $M = 0.5$, we are assuming that the unobserved confounding is half as much as the benchmarked confounding. Because the data is normalized, all quantities are on the scale of $[-1,1]$ and can be effectively interpreted as partial correlations. 

\begin{figure}[!h]
    \centering
    \caption{Contour Plots Across an Array of Exclusion Restriction Violation Scenarios}\par
    \includegraphics[scale=0.32]{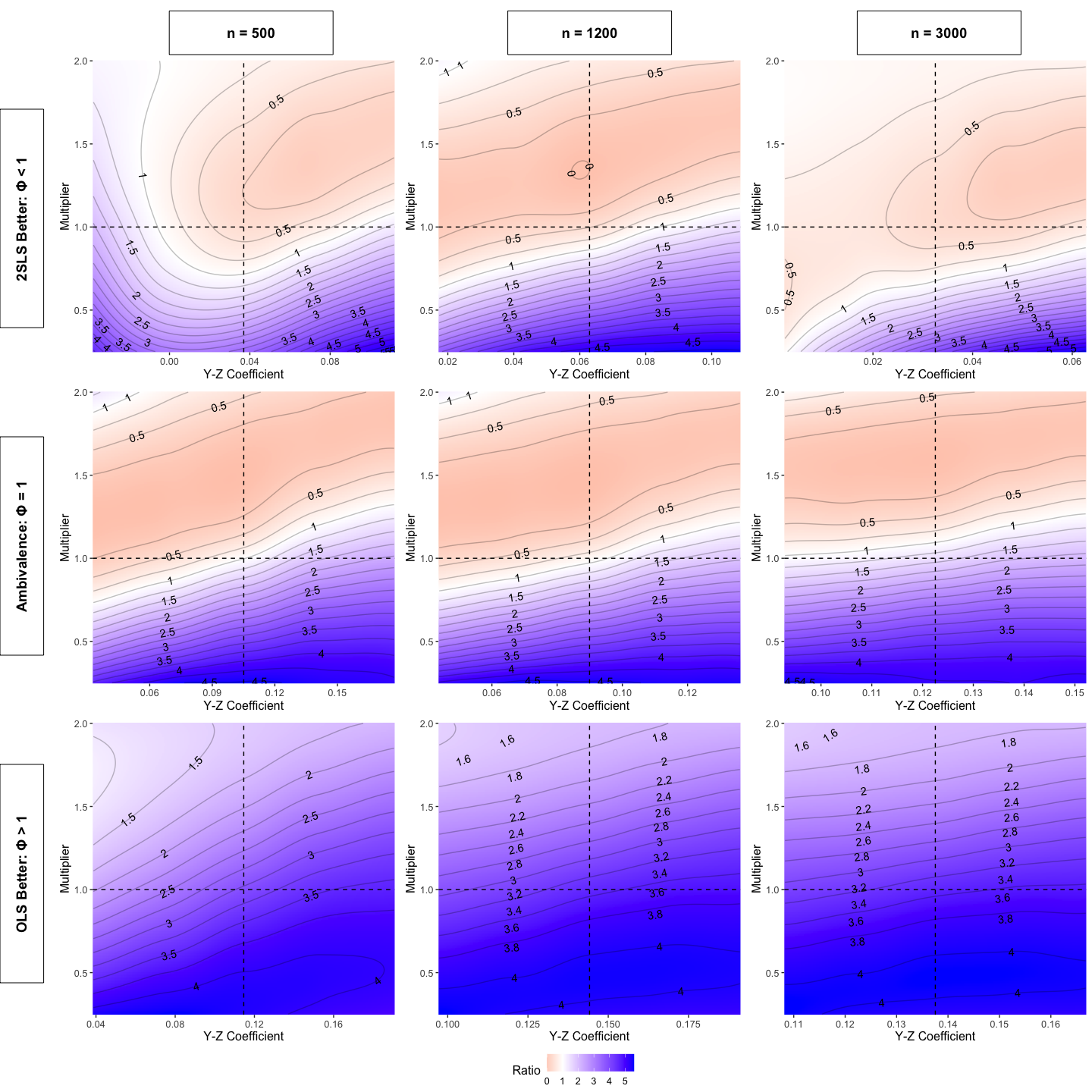}
    \label{fig:contour_er}
\end{figure}

With the ability to calculate an estimate on $\phi$, we can construct a graphical depiction of the relative trade-offs across a matrix of cases. That is, we can can calculate $\phi$ across an array of multipliers to account for several benchmark scenarios and, furthermore, across, for instance, the 90\% confidence interval for $\beta^R_{ER}$ to account for sampling variability in the estimation of the exclusion restriction violation. The final result is a contour plot colored by $\phi$ such as the one presented in Figure \ref{fig:contour_er}, which depicts the contour plots across a grid of example simulated scenarios where $\phi < 1$ (2SLS better), $\phi = 1$ (ambivalence), and $\phi > 1$ (OLS better) with sample sizes $n = 500, 1200, \text{and } 3000$ to show the impact of a widening confidence interval bounds on the decision-making. 

Focusing on the plot in the upper-right, the dotted lines represent for the x-axis and y-axis the point estimate for $\beta^R_{ER}$ and the multiplier of 1, respectively. The value of $\phi$ at the intersection of these dotted line indicates our decision if we take our data at face-value. The value of $\phi$ is about 0.5, indicating the inconsistency for the ACE of 2SLS due to an exclusion restriction violation is about half the inconsistency in OLS due to unmeasured confounding. This indicates that 2SLS may be the more appropriate method compared to OLS, which matches with our simulation. At a glance, referring to the legend, where the plot remains red (as opposed to white or blue) we maintain this conclusion. Nevertheless, we can see that as we move towards the upperbound of the confidence interval we need an increasingly larger multiplier on unmeasured confounding to cross into making the opposite conclusion due to $\phi > 1$. In the context of the current data analysis at hand, an analyst should consider the likelihood of both $\beta^R_{ER}$, which represents a partial correlation, nearly doubling and that the degree of unmeasured confounding is less than the available covariates to benchmark as it pertains to a certain multiplier.

As we move to the second row, ambivalence to methodology, we can see the intersection of the dotted lines lands on the white strip with translates to $\phi = 1$. Furthermore, roughly half the scenarios are colored red while the other half is blue further indicating graphically that there is no clear choice of 2SLS or OLS. In this case, our sensitivity analysis for the exclusion restriction assumption is inconclusive and that other assumptions and factors regarding the analysis should be examined. In the third row, it is clear that across virtually all the scenarios observed, the exclusion restriction violation is large enough such that the 2SLS has a higher inconsistency for the ACE compare to the OLS, sometimes at a ratio of nearly 4 times.

\subsection{Independence Violation}

For the independence assumption, our ratio is now
\begin{align*}
    \phi = \Big| \frac{c_2c_I}{c_3+\frac{c_1c_I}{1-c_7^2}}\Big|\Big|\frac{\frac{c_1c_2(1-c_7^2-c_I^2)}{1-c_7^2}}{1-(c_5+c_3c_7)^2-(1-c_7^2)(c_3+\frac{c_1c_I}{1-c_7^2})^2}\Big|^{-1}.
\end{align*}
Involving partial $R^2$s for this assumption is more tractable to calculate and interpretable in this setting as opposed to the coefficients themselves. Furthermore, the key edge weight $c_{I}$ plays a key role in both the inconsistency of 2SLS and OLS. Combining these both leads to a quantification of the degree of violation of the independence assumption using the partial Cohen's $f^2$ where $f^2_{U \sim Z|W} = \frac{R^2_{U \sim Z | W}}{1- R^2_{U \sim Z | W}}$. The partial Cohen's $f^2$ is a common measure for the general effect size of a relationship.\citep{cinelli_making_2020} The notation in the super-script indicates we are interested in the relationship between $U$ and $Z$ conditioning on $W$. The inconsistency ratio can be translated to
\begin{align*}
    \phi = f^2_{U \sim Z|W}\left(\frac{\sqrt{R^2_{X \sim U | Z,W}}sd(Z^{\perp W})sd(X^{\perp Z,W})R^2_{X \sim Z|W}\frac{Var(X^{\perp W})}{Var(Z^{\perp W})}}{1-R^2_{X \sim W}-(1-R^2_{X \sim W})(R^2_{X \sim Z|W}\frac{Var(X^{\perp W})}{Var(Z^{\perp W})})}\right)^{-2}
\end{align*}
where, for example, $sd(Z^{\perp W})$ represents the standard deviation of the residuals produced from regressing $Z$ on $W$ (see appendix for the construction of this result). 

\begin{figure}[!htb]
    \centering
    \caption{Contour Plots Across an Array of Independence Violation Scenarios}\par
    \includegraphics[scale=0.32]{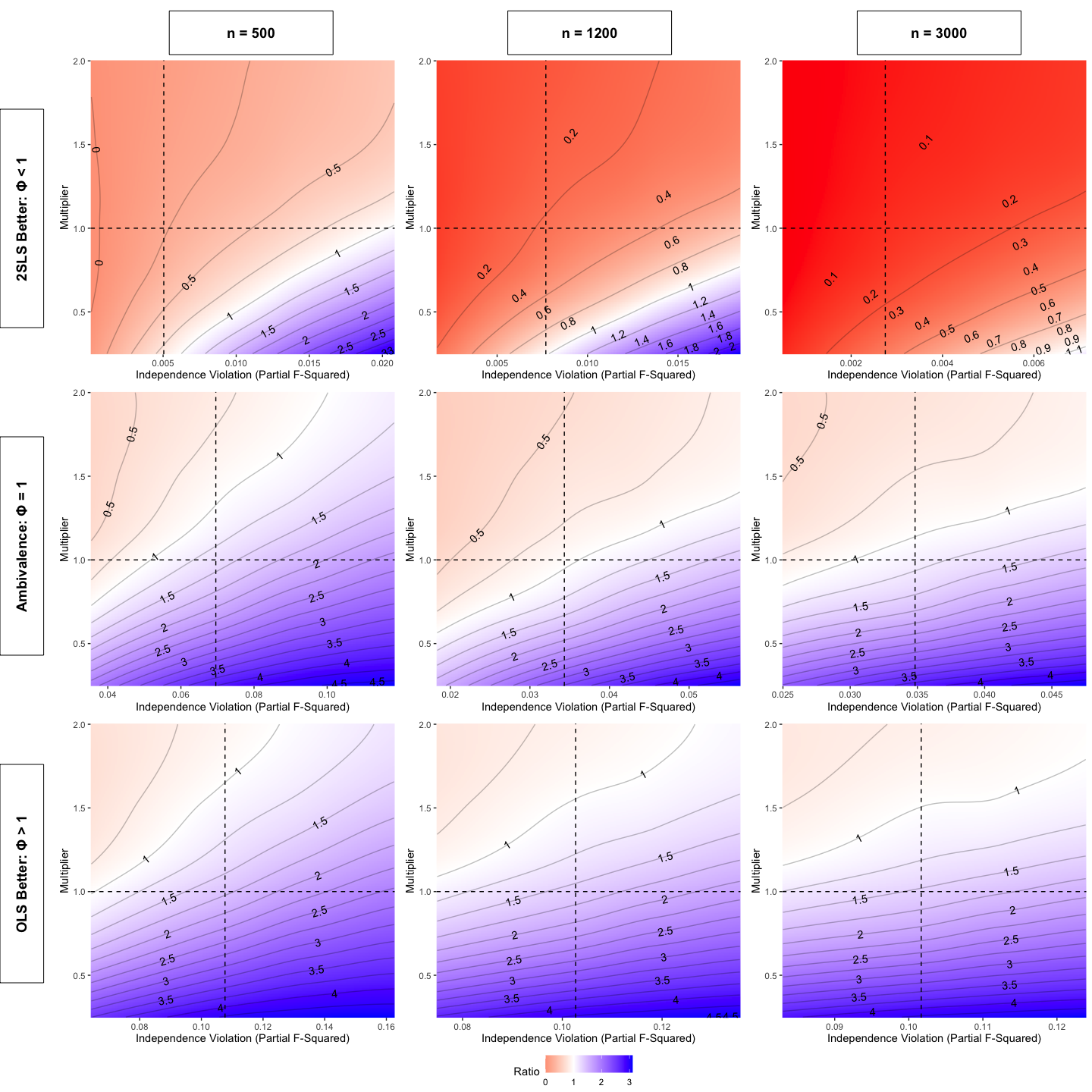}
    \label{fig:contour_ind}
\end{figure}

There are only two quantities we cannot estimate directly: $R^2_{U \sim Z | W}$, the degree of violation, and $R^2_{X \sim U|Z,W}$, the degree of unmeasured confounding that influences $X$. We will benchmark the former with $R^2_{Z \sim W_j|\mathcal{W}_{-j}}$ and the latter with $\max_j R^2_{X \sim W_j| \mathcal{W}_{-j},Z}$. Similar to the exclusion restriction, we will compute a 90\% confidence interval of the partial $f^2$ quantity (constructed via the bootstrap) and place a multiplier on the benchmarked unmeasured confounding. Using the same method to form the graphical representations, the contour plots across an array of scenarios are presented in Figure \ref{fig:contour_ind}. The graphs have the same interpretation only that the x-axis directly measures the degree of violation. For example, the top left plot, at face value, the dotted lines intersect at a $\phi$ of 0.25 meaning that the inconsistency in 2SLS due to a potential violation is four times smaller than that of OLS due to unmeasured confounding. Thus, we favor 2SLS in this case. Where the white ambivalence line intersects the horizontal dotted line tells us that if we trust our benchmark at a multiplier of one, then the effect size will need to increase to about 0.02 to be ambivalent. The feasibility of this can be judged using subject matter expertise in a given analysis.

\subsection{Treatment Effect Heterogeneity}

As previously discussed, we will be comparing the inconsistency of OLS that does not adjust for $Z$ (i.e. $\lambda_2$) to the estimate produced under 2SLS. We have that
\begin{align*}
    \phi = \Bigg|\frac{\alpha_1\alpha_3\beta_3 - \frac{2\alpha_1\alpha_3\alpha_5^2\beta_3}{\alpha_1^2 + \alpha_5^2}}{\alpha_1^2 + \alpha_5^2 - \frac{4\alpha_1^2\alpha_5^2}{(\alpha_1^2 + \alpha_5^2})}\Bigg|\Bigg|\frac{\alpha_2\beta_2 + 2\alpha_1\alpha_3\beta_3}{1-\alpha_4^2-\frac{4\alpha_1^2\alpha_5^2}{1 + \alpha_4^2 + 2\alpha_5^2}}\Bigg|^{-1}.
\end{align*}
\noindent The key quantity to measure the heterogeneity violation is $\alpha_3$, or the coefficient representing the $Z-U$ interaction. After re-arranging terms (see appendix), we can re-express the inconsistency ratio as
\begin{align*}
    \phi = |\alpha_3|\Bigg(\Bigg|\frac{(\alpha_1^2 - \alpha_5^2)^2}{(\frac{1}{2}\alpha_1^2 - \frac{3}{2}\alpha_5^2)Var(X^{\perp W, XW})- (\alpha_1^2 - \alpha_5^2)^2}\Bigg|\Bigg|\frac{\beta_2\alpha_2}{2\alpha_1\beta_3}\Bigg|\Bigg)^{-1}
\end{align*}

We can benchmark $\alpha_3$ by via the regression coefficient that the corresponds to the maximum absolute value of the coefficient between the observed covariates and $Z$ in the model $E[X|z,w_1,w_2,...,w_J,zw_1,zw_2,..zw_J]$. Because this value comes directly from the regression model, we can easily derive 90\% confidence intervals. $\beta_3$, or the $X-U$ interaction can be benchmarked in a similar fashion. Similar to $c_1c_2$ in the exclusion restriction and because all variables are centered, $\alpha_2\beta_2$ measures the degree of marginal unmeasured confounding, which we can benchmark in the same manner and include multipliers on these quantities. The only caveat is that we need to make sure to include the interaction terms in the benchmark model. 

\begin{figure}[!htb]
    \centering
    \caption{Contour Plots Across an Array of Treatment Effect Heterogeneity Scenarios}\par
    \includegraphics[scale=0.32]{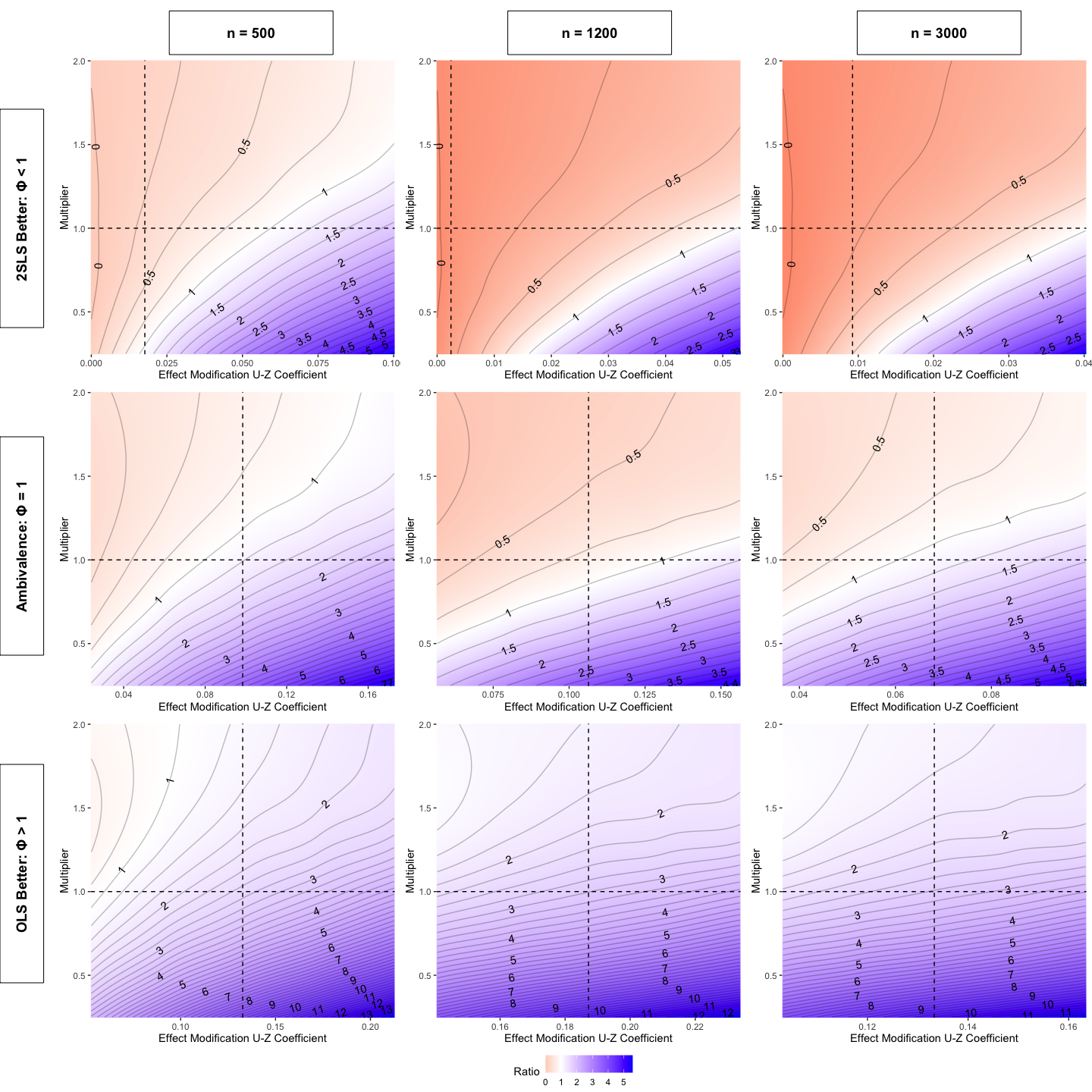}
    \label{fig:contour_het}
\end{figure}

Figure \ref{fig:contour_het} presents the contour plots across several scenarios where on the x-axis the absolute value $\alpha_3$ is estimated at the vertical dotted line. Furthermore, the lowest and highest absolute values of the bounds of the confidence interval are plotted as the range in the plot. The bottom left plot shows that there is significant heterogeneity, obtaining a coefficient of 0.135 on the partial correlation scale, which produces a $\phi$ of about 2.5. This means that the heterogeneity pertaining to $Z$ is large enough such that the LACE is 3 times more inconsistent for the ACE than OLS despite unmeasured confounding. If we are at the lower end of the range and if the true degree of unmeasured confounding is twice that of our benchmark, then our $\phi$ is closer to 1 and we are more ambivalent in regards to this assumption violation causing inconsistency greater than that of OLS.

\section{Simulation Results}

In this section, we present results verifying the closed form derivations for all three assumption violation scenarios with and without covariates. Additionally, we further present results validating the accuracy of $\phi$ to capture the inconsistency. All variables were generated via the structural equations provided in the earlier sections with the error terms following a normal distribution with mean zero. When the variables were independent from all other variables in the system, such as $U$, they were standard normal. When the variable was determined by other variables in the system, such as $X$, it was still normal but with the variance of the error term being equal to one minus the variance of the other variables in the structural equation. This is so that the total variance of terms like $Var(X)$ will still equal one (see the "Notes about Simulations" section in the appendix). For example, in Eq \ref{eq:simple_first},  $\sigma_{\epsilon_1} = 1 - c_1^2 - c_3^2$ and, thus, $\epsilon_1 \sim N(0,1 - c_1^2 - c_3^2)$. 

To obtain the simulated numbers for OLS, we use the built-in \texttt{lm} from R function while for 2SLS, we use the \texttt{ivreg} function from the \texttt{ivreg} R package. In the exclusion restriction and independence setting, we present the Monte Carlo averages over 500 simulations of 500 observations generated. For treatment effect heterogeneity because the empirical results take more samples to converge, we used 500 simulations of 3000 observations. Note that for all simulations, for demonstration purposes, we set the IV to be sufficiently strong such that the estimates would converge on the population value within a reasonable sample size. Nevertheless, our results otherwise hold for weak IVs if the number of observations in each simulation increased significantly. 

To demonstrate the sensitivity analysis plots, we examine an array of scenarios that scale the degree of violation from zero to a value where the inconsistency of 2SLS is far past that of OLS. At each degree of violation, we generate 500 samples from the data generating mechanism of sizes $n = 500, 1200,\text{and} 3000$ and record the proportion of calculated $\phi$'s that are above 1 given a multiplier of $1$. As the violation increases, this proportion should increase from $0$ to $1$. Crucially, it should also equal $0.5$ when we reach ambivalence between OLS and 2SLS. To simulate a set of covariates that we may use to benchmark the unobserved quantities, we generated three independent covariates ($W_1,W_2,W_3$) from a standard normal distribution and constructed $W$ via the first principal component.

\subsection{Exclusion Restriction}

For the case with no covariates, we set the following structural parameters $c_0 = 0.3$, $c_1 = 0.5$, $c_2 = 0.5$, $c_3 = 0.5$, and $c_{ER} = 0.25$. For the case with covariates, we have $c_0 = 0.25$, $c_1 = 0.4$, $c_2 = 0.4$, $c_3 = 0.7$, $c_{ER} = 0.25$, $c_5 = 0.4$, and $c_6 = 0.4$. The results are presented in Table \ref{tab:er_sim}.

\begin{table}[!htbp]
  \centering
  \caption{Theoretical and Simulated Results for the Exclusion Restriction Violation}
  \begin{tabular}{ccccc}
    \toprule
    & & \multicolumn{3}{c}{Method} \\
    \cmidrule(lr){3-5}
    & & OLS without Z & OLS with Z & 2SLS with Z \\
    \midrule
    \multirow{2}{*}{Without covariates} & Closed Form Result & 0.375 & 0.333 & 0.5 \\
    & Simulated Result & 0.374 & 0.334 & 0.496 \\
    \midrule
    \multirow{2}{*}{With covariates} & Closed Form Result & 0.399 & 0.457 & 0.357 \\
    & Simulated Result & 0.398 & 0.459 & 0.354 \\
    \bottomrule
  \end{tabular}
  \label{tab:er_sim}
\end{table}

Using the values $c_1 = c_2 = 0.35$, $c_3 = 0.3$, and $c_5 = c_6 = 0.385$ (to account for $W$ being constructed via PCA for benchmarking), Figure \ref{fig:logit_plot_ind} demonstrates the properties of $\phi$ as the exclusion violation grows in magnitude. As expected, when the violation is low, the average $\phi$ begins low and steadily increases before leveling off near one. The vertical dotted line represents magnitude of $c_{ER}$ where the inconsistency of OLS is equal to that of 2SLS. For all sample sizes examined, we see that at this ambivalence point the average $\phi$ is about 0.5, indicating good performance of our procedure. Predictably, as the sample size increases, the behavior of $\phi$ at the ends of the range for the degrees violation (i.e. no violation and large) is more stable as there is less variability in the calculation of $\phi$.

\begin{figure}[!h]
    \centering
    \includegraphics[scale = 0.5]{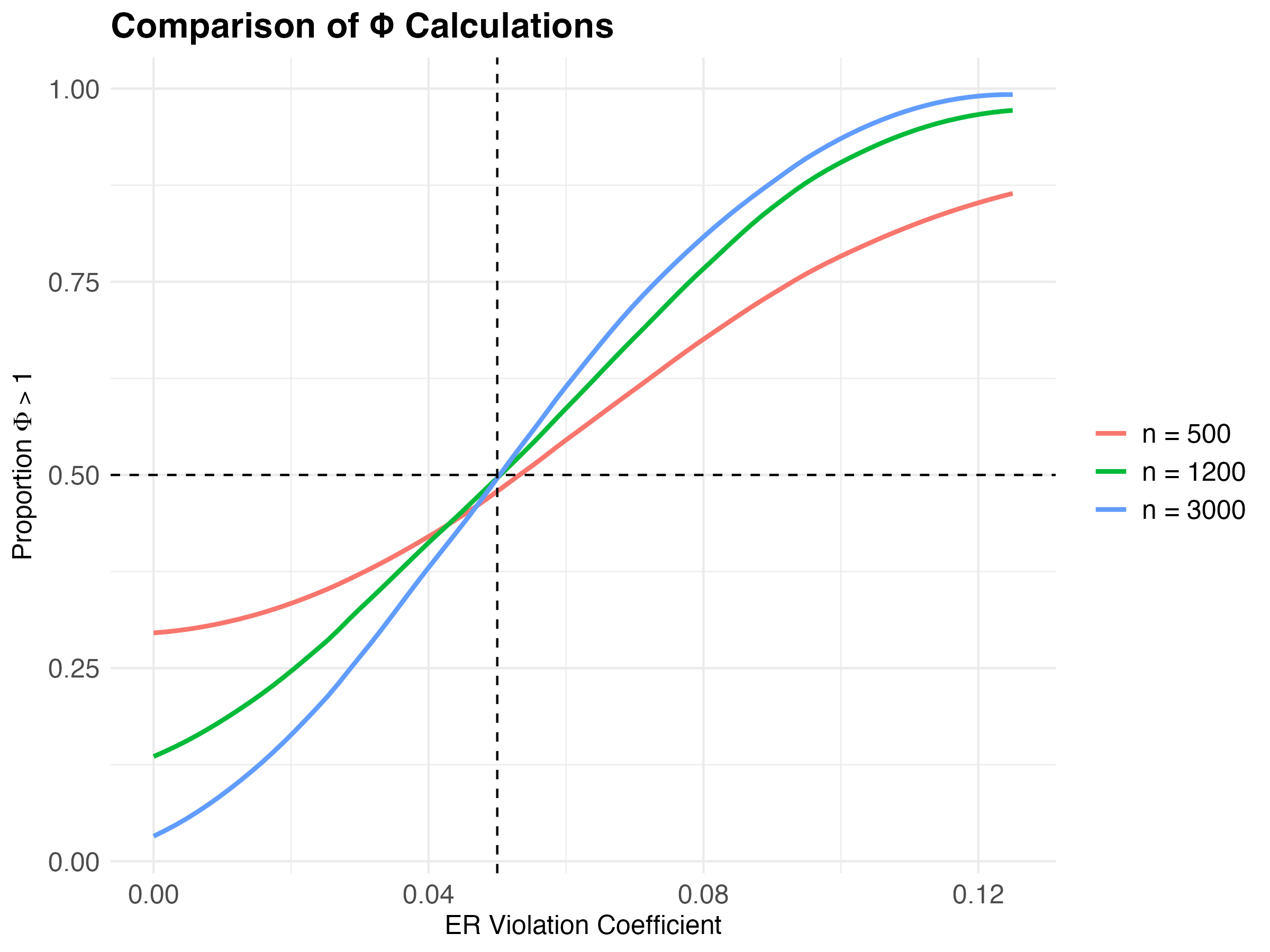}
    \label{fig:logit_plot_ind}
    \caption{Average $\phi$ across 500 simulations of the given sample size over a variety of degree of exclusion restriction violations. The vertical dotted line represents the point of ambivalence.}
\end{figure}

\subsection{Independence}

For the case with no covariates, we set the following structural parameters $c_0 = 0.3$, $c_1 = 0.5$, $c_2 = 0.5$, $c_3 = 0.5$, $c_{ER} = 0.25$. For the case with covariates, we have $c_0 = 0.3$, $c_1 = 0.4$, $c_2 = 0.4$, $c_3 = 0.5$, $c_{ER} = 0.25$, $c_5 = 0.4$, $c_6 = 0.4$, and $c_7 = 0.25$. The results are presented in Table \ref{tab:uc_sim}.

\begin{table}[!htbp]
  \centering
  \caption{Theoretical and Simulated Results for the Independence Violation}
  \begin{tabular}{ccccc}
    \toprule
    & & \multicolumn{3}{c}{Method} \\
    \cmidrule(lr){3-5}
    & & OLS without Z & OLS with Z & 2SLS with Z \\
    \midrule
    \multirow{2}{*}{Without covariates} & Closed Form Result & 0.312 & 0.384 & 0.2 \\
    & Simulated Result & 0.311 & 0.383 & 0.200 \\
    \midrule
    \multirow{2}{*}{With covariates} & Closed Form Result & 0.290 & 0.391 & 0.176 \\
    & Simulated Result & 0.290 & 0.394 & 0.178 \\
    \bottomrule
  \end{tabular}
  \label{tab:uc_sim}
\end{table}

In Figure \ref{fig:logit_plot_ind}, we present the results of increasing the independence violation with $c_1 = c_2 = 0.25$, $c_3 = 0.2$, and $c_5 = c_6 = 0.35$ and $c_7$ scaling with the degree of violation to ensure accurate benchmarking. Based on the plotted proportions, the performance of $\phi$ matches our expectations. 

\begin{figure}[!h]
    \centering
    \includegraphics[scale = 0.5]{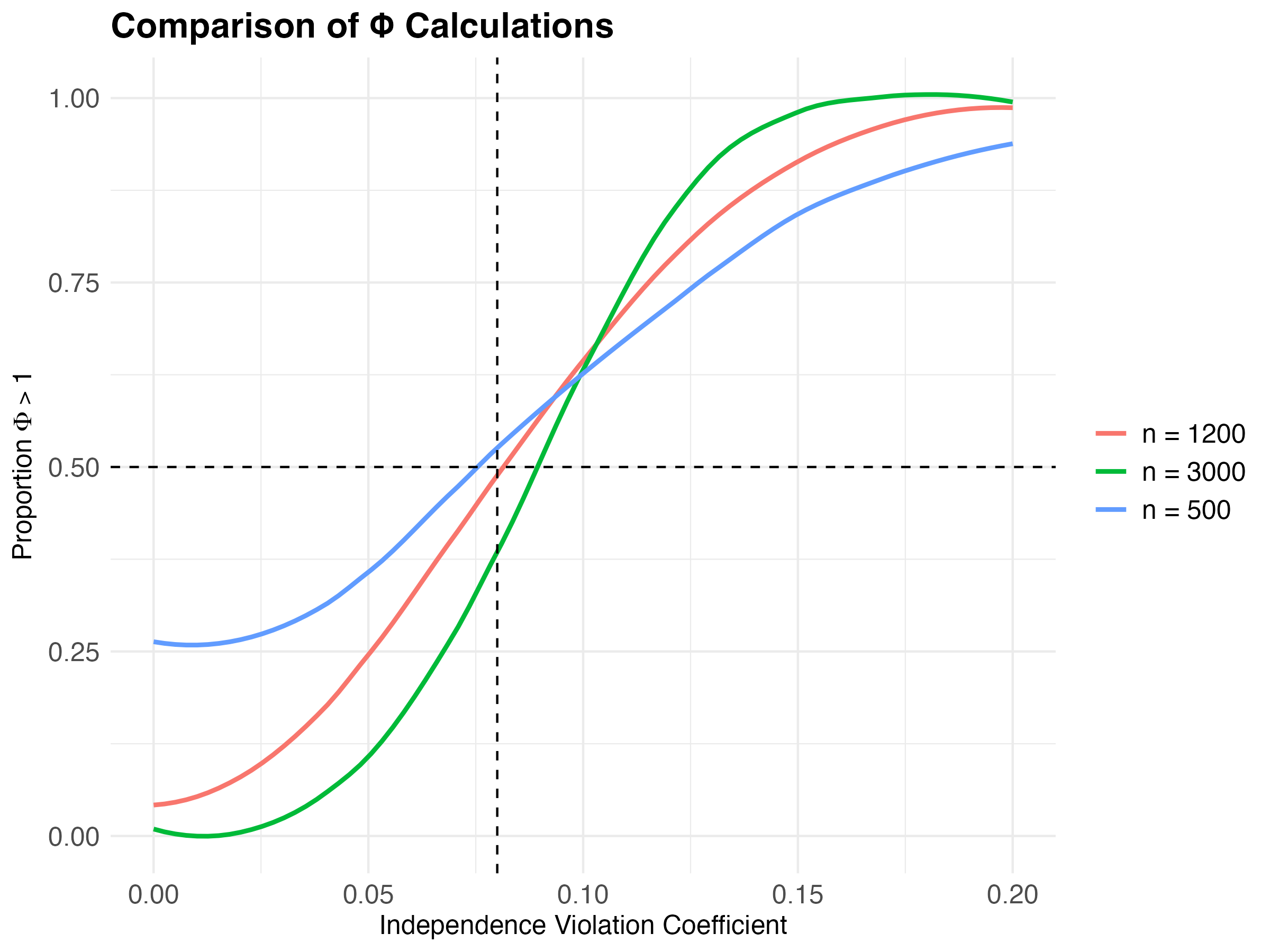}
    \caption{Average $\phi$ across 500 simulations of the given sample size over a variety of degrees of independence violations.}
    \label{fig:logit_plot_ind}
\end{figure}

\subsection{Treatment Effect Heterogeneity}

\begin{figure}[!h]
    \centering
    \includegraphics[scale = 0.5]{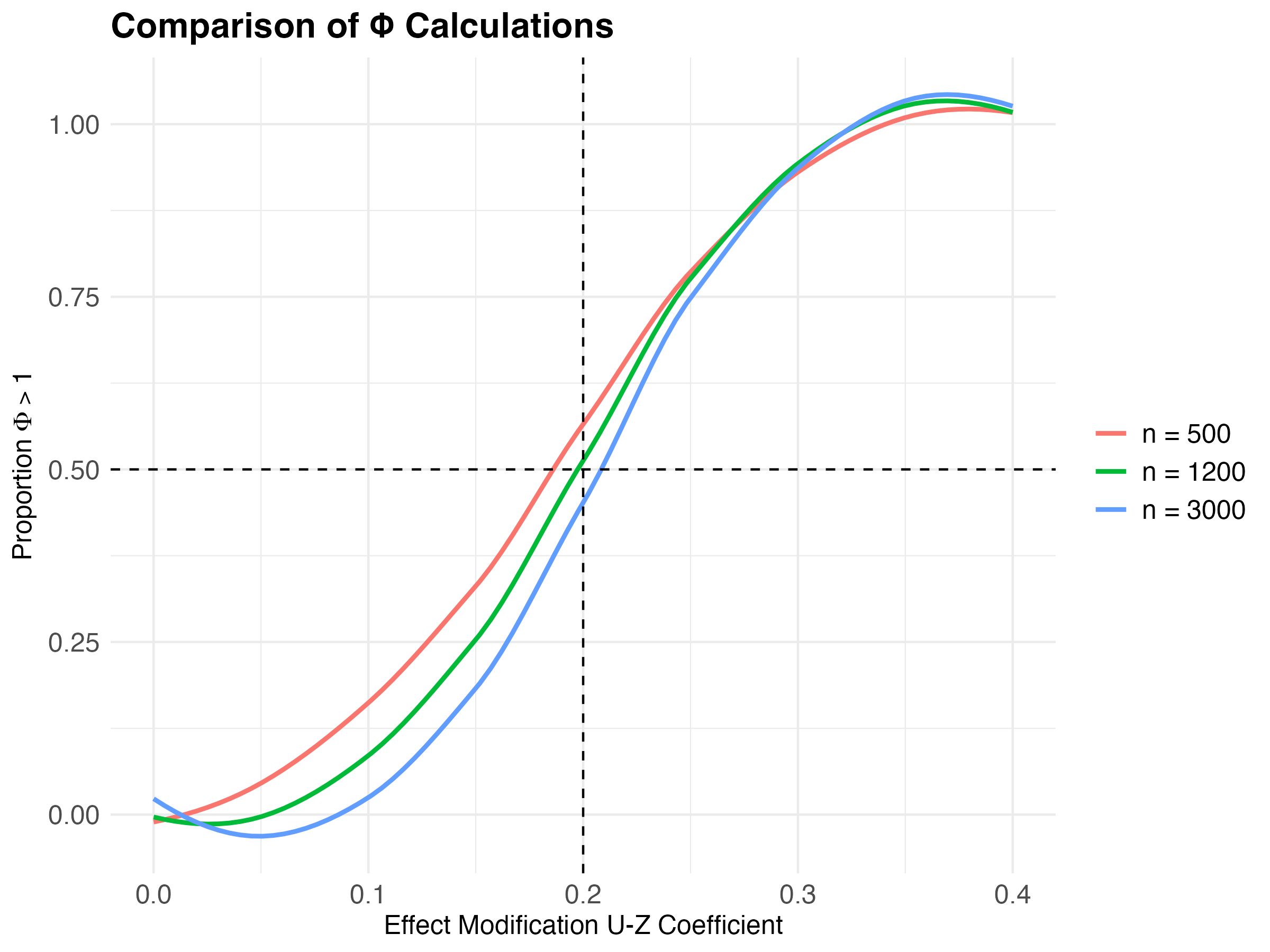}
    \caption{Average $\phi$ across 500 simulations of the given sample size over a variety of degrees of treatment effect heterogeneity interactions as measured by the effect modification via the interaction coefficient between $U$ and $Z$ in the first stage.}
    \label{fig:logit_plot_heterogeneity}
\end{figure}

For the case with no covariates, we set the following structural parameters $\beta_1 = 0.1$, $\beta_2 = 0.2$, $\beta_3 = 0.1$, $\alpha_1 = 0.45$, $\alpha_2 = 0.15$, and $\alpha_3 = 0.1$. For the case with covariates, we have $\beta_1 = 0.1$, $\beta_2 = 0.2$, $\beta_3 = 0.1$, $\beta_4$, $\beta_5$, $\alpha_1 = 0.45$, $\alpha_2 = 0.15$, $\alpha_3 = 0.1$, $\alpha_4 = 0.15$, and $\alpha_5 = 0.1$. The results are presented in Table \ref{tab:em_sim}. Figure \ref{fig:logit_plot_heterogeneity} shows the performance of $\phi$ when $\beta_2 = 0.25$, $\beta_3 = 0.15$, $\beta_4 = 0.35$, $\beta_5 = 0.15$, $\alpha_1 = 0.4$, $\alpha_2 = 0.25$, and $\alpha_4 = 0.25$.

\begin{table}[!htbp]
  \centering
  \caption{Theoretical and Simulated Results for Treatment Effect Heterogeneity}
  \begin{tabular}{ccccc}
    \toprule
    & & \multicolumn{3}{c}{Method} \\
    \cmidrule(lr){3-5}
    & & OLS without Z & OLS with Z & 2SLS with Z \\
    \midrule
    \multirow{2}{*}{Without covariates} & Closed Form Result & 0.039 & 0.044 & 0.022 \\
    & Simulated Result & 0.039 & 0.044 & 0.021 \\
    \midrule
    \multirow{2}{*}{With covariates} & Closed Form Result & 0.040 & 0.037 & 0.021 \\
    & Simulated Result & 0.039 & 0.046 & 0.021\\
    \bottomrule
  \end{tabular}
  \label{tab:em_sim}
\end{table}

\section{Applied Example}

The causal effect of educational achievement on earnings have been the subject of several observational studies including the widely-cited results of Card (1993), which studies a sample of $n = 3,010$ men from the National Longitudinal Survey of Young Men (NLSYM).\cite{card_using_1993} More details and access to the data can be found in \texttt{R} via \texttt{wooldridge} package under the \texttt{card} object. After log transforming earnings, Card (1993) found via OLS that there was a statistically significant 7.5\% increase in earnings for each year of education. Nevertheless, there is large potential for residual confounding despite adjusting for race, experience, and regional indicators. As such, Card (1993) pursued an IV approach using an indicator variable for if the participant lived near a four-year college during their teenage years or not, which we will refer to as "Proximity." With this approach, 2SLS finds a significant 13.2\% increase though with larger confidence intervals compared to OLS. Even so, as discussed in Cinelli and Hazlett (2022) there is concern that this IV may be invalid due to unmeasured geographical factors that are potentially associated with Proximity and Earnings.\citep{cinelli_omitted_2022}

\begin{figure}[h!]
    \centering
    \caption{Examining the Three Assumptions for the Proximity IV}\par
    \includegraphics[scale=0.2]{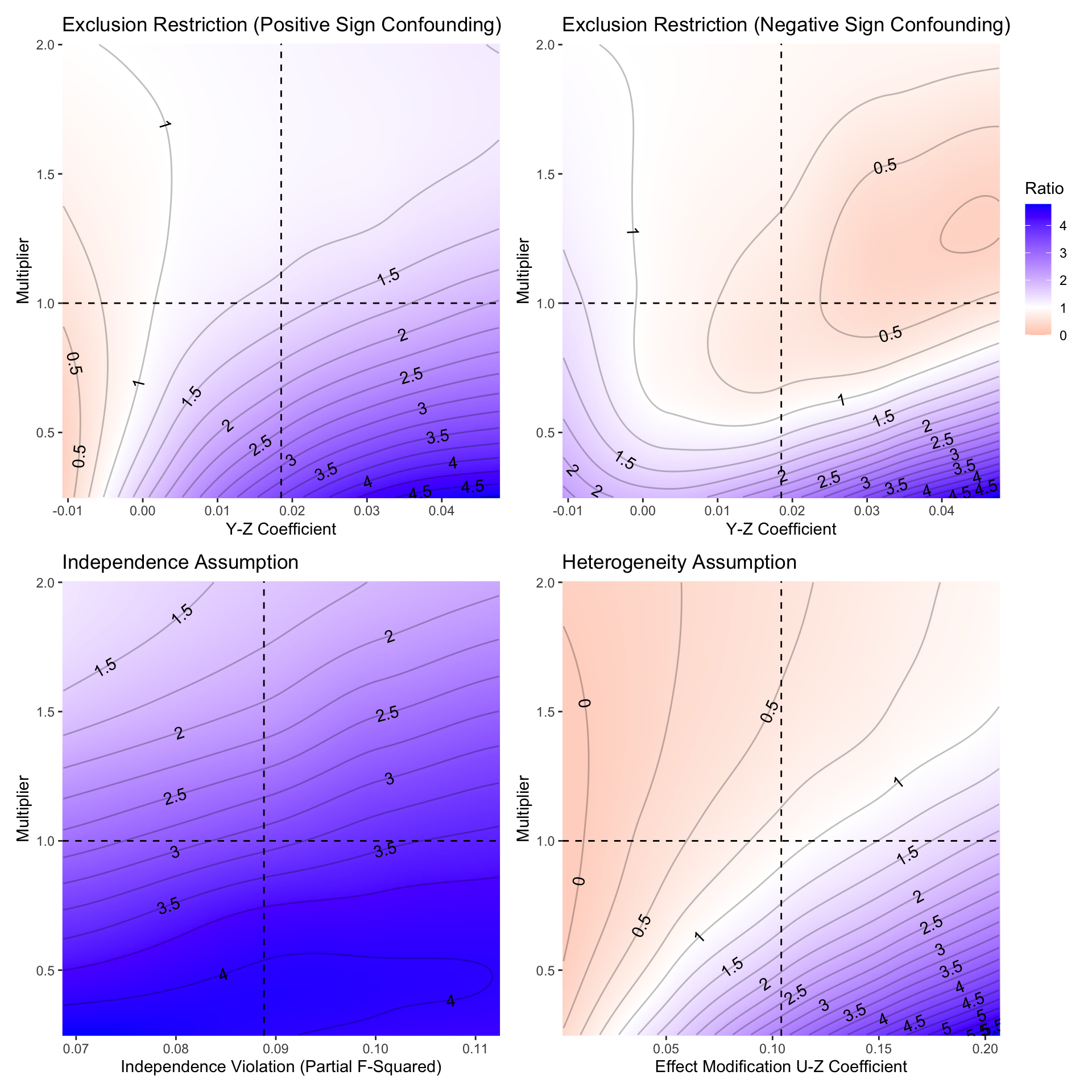}
    \label{fig:proximity_sens}
\end{figure}

Though we use the same application, our sensitivity analysis addresses a different question than Cinelli and Hazlett (2022). Where as they aim to measure how large the strength of an an omitted variable that influences the IV and outcome must be to render 2SLS results null, we instead inquire whether for each assumption there exists a violation large enough to result in worse inconsistency than OLS. The findings of our procedure are presented in Figure \ref{fig:proximity_sens}. For sake of example, we interpret our findings based on the intersection of the dotted lines for each plot. The vertical dotted line represents the benchmarked assumption violation based on the data while the horizontal line denotes the multiplier of one, which assumes our data roughly quantifies the degree of unmeasured confounding.

With no prior knowledge on the sign of confounding, we may choose to average the results from the two exclusion restriction plots where $\phi$ takes the value of about 1.4 and 0.65 for the positive and negative sign respectively. This leads us to be ambivalent on whether the direct relationship between proximity and earnings is large enough such that the 2SLS estimate is more or less inconsistent than the OLS estimate. For heterogeneity, we have benchmarked a notable degree of interaction between an unmeasured confounder and proximity in the first stage, leading to a value of $\phi$ of approximately 0.9, slightly favoring 2SLS in this case. That is, there is likely some inconsistency in the LACE for the ACE but not to the amount where it renders the 2SLS estimate more inconsistent than the OLS estimate. Therefore, the deciding factor is in the independence assumption, which shows a rather large violation at a $\phi$ of nearly 3.25. This means that an unmeasured confounder with a partial $f^2$ of 0.09 has rendered the inconsistency of 2SLS to be 3.25 times larger than that of OLS. We thus conclude that using proximity as an instrument in 2SLS is sub-optimal compared to OLS due to a large potential violation in the independence assumption – a similar conclusion to that of Cinelli and Hazlett (2022).

\section{Discussion}

In this work, we have investigated two predominant ways that analysts may isolate causal effects: CAC and IVAC via OLS and 2SLS, respectively. Considering these paradigms, we have based our study on the notion that each approach may work imperfectly due to assumption violations (as is most plausible in the vast majority of real world settings). Our closed-form results that capture interpretable rules of thumb based on DAG edge weights and coefficients, as well as our sensitivity analysis procedure, help guide analysts towards a practical philosophy on how one may execute observational studies. If the goal is to obtain an estimate of the ACE and there is a set of tools that one must choose from, such as the CAC and IVAC. Thus, one must ask: when is one tool more advantageous than the other? We have ultimately broken down this question by juxtaposing the degree of unobserved confounding to the degree IV assumption violations, providing results for analysts to offer evidence that the estimate computed was the least inconsistent it could have been given the scenario.

Our results suggest that properly defining confounders and IVs in the study paradigm is only a starting point. Whether a variable will be used in CAC or as an IV in IVAC is not necessarily congruent with its formal definition. In fact, we have presented analytically that, relative to OLS, there are scenarios where a variable should be used as an IV even though it does not meet the strict definition of an IV. For example, in an exclusion restriction violation the variable used as an IV is, by definition, a confounder and, although such a variable is not a "perfect IV," the relative performance of 2SLS would be better than that of OLS. Our sensitivity analysis procedure assists in detecting this by using the observed data.

Another relevant scenario lies in the fact that even though we may have a valid IV, the LACE estimand from 2SLS may be further away from the ACE than an estimand from OLS that is impacted by unobserved confounding. Through our closed form results and sensitivity analysis, we allow an analyst to judge how large heterogeneity would need to be in order for this to occur. Whether the resulting IV estimand remains scientifically useful is a subject of debate that we will not discuss here.\citep{imbens_better_2010} Rather, we are interested in directly targeting the ACE and will assume that treatment effect heterogeneity may provide a barrier to this goal. In this sense, the results from our paper may provide case-by-case evidence for and against those who may argue IV analyses may still be useful for the ACE.

In our sensitivity analysis tool, we provide separate evaluations for the three distinct IV assumption violation scenarios. If all evaluations agree that either 2SLS is superior or OLS is superior then the suggested approach is clear to the analyst. However, one may encounter a scenario that the graphs disagree: for example, the exclusion restriction and independence show 2SLS is superior while the treatment effect heterogeneity graph does not. In this case, one should rank the importance of each assumption and consider the observed benchmarks for each assumption violation. Using the multipliers provided or inputting a custom multiplier, one could consider the degree of unobserved confounding across the board in order for all three graphs to agree and evaluate whether these findings are reasonable in the particular study scenario. Similarly, one may further use the implied assumption violations in the legend relative to the benchmarked assumption violations for these purposes. Ultimately, sensitivity analyses are matters of judgement and we aim to provide quantitative tools for analysts to navigate these scenarios.

Though we focus on inconsistency in this paper, in practice, precision is important to consider as well. Even so, we believe that the first, and most important step, in any analysis is to verify that the correct estimand is being targeted, which requires reasoning about untestable causal assumptions. Our methods aim to assist analysts with this step. In the second step, the analyst will then judge which estimand is most efficient. This can be approximated by using the observed data to compare the standard error of each method. For instance, it would be clear that with a weak IV the confidence intervals will be comparatively wide. Of course, under unobserved confounding, there are implications on the first and second moment and, thus, inference. Examining the relative impacts on inference is a potential avenue of future research and could possibly be based upon the results of Cinelli and Hazlett.\citep{cinelli_making_2020,cinelli_omitted_2022}

Users of CAC often opt to capture confounding by adjusting for the propensity score (PS) using the fitted values of the exposure regressed on the confounders. Specifically, in the case of a continuous treatment, one will utilize the generalized propensity score (GPS).\citep{imbens_propensity_2004,imai_causal_2004} Like OLS, propensity scores estimates require conditional ignorability to be consistent so it falls in the purview of our scenarios (i.e. to adjust for $Z$ or not). There are several ways to utilize the propensity score including matching, stratification, adjustment, and weighting. If the functional form of the propensity score is correct, then both the adjustment and propensity score methods would be consistent. In the case of unmeasured confounding, the inconsistency of each method would not be notably different because the propensity score would be incorrect by the same degree. Subsequently, in terms of judging consistency, the results we present for OLS would be identical to propensity score via adjustment for the propensity score as opposed to individual covariates. Because we encapsulate all confounders in a single covariate $W$, we could replace $W$ and $Z$ with a GPS containing a linear specification and achieve the same results. Though, as a result, the effect of $Z$ in OLS would be mediated through the GPS, which would make our closed form derivations more complicated.

The results in this paper also have implications on more sophisticated non-parametric, data-driven variable selection techniques. In order to reduce Type I error, one would model the PS away from the outcome and, thus, it appears reasonable to use penalization or machine learning (ML) techniques to optimize prediction error. There are two main issues with this automated procedure: (1) strong though imperfect IVs would be selected leading to possible bias amplification and (2) omitted variable bias may occur due to shrinkage to zero of important confounders. These points have been mentioned by others\citep{pearl_class_2012,bhattacharya_instrumental_2007,chernozhukov_post-selection_2015} and our results further provide analysts information to act in the face of such issues. As a starting point, one may use the general intuition gleaned from our results to quantify how strong the confounder or IV should be in order to avoid adjustment altogether or, instead, use it in an IVAC framework. Subsequently, future directions may include extending the framework developed in this paper to techniques such as augmented inverse probability of treatment weighting, post-LASSO, targeted minimum loss estimation, and double machine learning to weigh the approaches against one another.\citep{van_der_laan_targeted_2010,belloni_sparse_2012,chernozhukov_doubledebiased_2018}

There are several additional avenues for future research. As previously mentioned, we focus only on consistency but in estimation, we may also want to know whether CAC may be more efficient than IVAC or vice-versa. Additionally, confidence interval overlap of OLS and 2SLS estimates could also cause ambivalence between the methods. Another future direction lies in moving beyond the setting of a continuous exposure and outcome. Nevertheless, if one believes linear probability models (LPMs) are appropriate for the study's context, then one may extend our results to a binary exposure and outcome. It has been shown that if the probabilities produced by the LPM are not outside of the range of $[0,1]$, or that the probabilities of exposure or outcome are not extreme in the population, then OLS and 2SLS may still give consistent results.\citep{horrace_results_2006,basu_2sls_2018} For the sensitivity analysis procedure, the results are only as useful as variables available and chosen for benchmarking, which is partially mitigated by using the multipliers. Certainly, other benchmarking quantities for the unobserved quantities than the ones we chose could be used. The aggregation of the covariates into one general confounder $W$ could be done via other methods besides PCA, which is limited when there are categorical variables. Using non-continuous variables is also limited when capturing associations using $R^2$ due to the Frechet bounds on the correlation between non-continuous variables being potentially far narrower than $[-1,1]$.\citep{horn_matrix_1985}

Isolating causal effects in observational data presents many challenges, which foremost include the effect of unobserved confounding. The CAC and IVAC offer potential avenues to mitigate this confounding. Even so, under untestable assumptions, choosing the optimal approach for the problem at hand involves much conjecture. Our closed-form findings and sensitivity analysis approach helps analysts quantitatively justify the approach that they ultimately believe produces an estimate closest to the ACE. The upshot is that, with the information we provide, results from observational studies will both be more transparent and more useful in their interpretation.

\section{Acknowledgements}

RSZ and DLG are supported by NIH/NIA P30AG066519 and NIH/NIA 1RF1AG075107.

\section{Bibliography}

\bibliographystyle{unsrtnat}
\bibliography{references}

\section{Appendix}

\subsection{Proof for the Consistency $\hat{A}_4$ under Figure \ref{fig:dag_simple}}
\begin{proof}
We begin by defining following system of structural equations under Figure \ref{fig:dag_simple} where $E[\epsilon_1|U,Z] = 0$ and $E[\epsilon_2|X,U] = 0$. 
\begin{equation}
    X = \alpha_1U + \alpha_2Z + \epsilon_1
\end{equation}
\begin{equation}
    Y = \beta_1X + \beta_2U + \epsilon_2
\end{equation}

By the standard definition of the IV estimator, $\hat{A}_4 = \frac{\widehat{Cov}(Y,Z)}{\widehat{Cov}(X,Z)}$. Where $n$ is the sample size, by finite variance and Slutsky's theorem we can write:
\begin{align}
    \frac{\widehat{Cov}(Y,Z)}{\widehat{Cov}(X,Z)} &= \frac{n^{-1}\sum_{i=1}^{n}y_iz_i}{n^{-1}\sum_{i=1}^{n}x_iz_i} \overset{p}{\to}\frac{E[YZ]}{E[XZ]} = \frac{Cov(Y,Z)}{Cov(X,Z)}
\end{align}

Clearly, $Cov(X,Z) = \alpha_2 = c_3$ because $Z \ind U$. Moving onto $Cov(Y,Z)$ we write:

\begin{align}
    E[Y|z] &= \sum_{u}E[Y|z,u]P(u|z)\\
    &= \sum_{u}[c_0c_3z + c_2u]P(u)\\ 
    &= c_0c_3z\sum_{u}u + c_2\sum_{u}uP(u)\\
    &= c_0 c_3z
\end{align}
\end{proof}

\subsection{Proof for Proposition \ref{rmk:er}}
\begin{proof}

Using an equivalent definition of the IV estimator, $\hat{A}_4 = (Z^T X)^{-1}(Z^TY)$ we can substitute \ref{eq:er_eq2} for $Y$

\begin{align}
    (Z^T X)^{-1}(Z^TY) &= (Z^T X)^{-1}(Z^T(c_0X + c_2U + c_{ER}Z + \epsilon_3))\\
    &= c_0 + c_2(Z^T X)^{-1}Z^TU + c_{ER}(Z^T X)^{-1}Z^TZ + (Z^T X)^{-1}Z^T\epsilon_3\\
    &= c_0 + c_2\frac{n^{-1}\sum_{i=1}^{n}z_iu_i}{n^{-1}\sum_{i=1}^{n}z_ix_i} + c_{ER}\frac{n^{-1}\sum_{i=1}^{n}z_i^2}{n^{-1}\sum_{i=1}^{n}z_ix_i} + \frac{n^{-1}\sum_{i=1}^{n}\epsilon_{3i}x_i}{n^{-1}\sum_{i=1}^{n}z_ix_i}\\
    &\overset{p}{\to} c_0 + c_2\frac{Cov(Z,U)}{Cov(Z,X)} + c_{ER}\frac{Var(Z)}{Cov(Z,X)} + \frac{Cov(Z,\epsilon_3)}{Cov(Z,X}\\
    &= c_0 + \frac{c_{ER}}{c_3} & \text{($Z \ind U$ and $Cov(X,Z) = c_3$)}
\end{align}
The last line follows from the fact that $Z \ind U$ and $E[\epsilon_3Z] = 0$ by iterated expectation.
\end{proof}

\subsection{Proofs for Proposition \ref{rmk:uc}}
\subsubsection{Proof for the Consistency of $\hat{A}_3$}
\begin{proof}
We need to find the value of $\frac{\partial}{\partial x}E[Y|x,z] = c_0 + c_2\frac{\partial}{\partial x}E[U|x,z]$. We will use the FWL theorem to find $\frac{\partial}{\partial x}E[U|x,z]$ by orthogonalizing $Z$ and using consistency. Assuming no intercept, we can write the quantity as $\frac{X^TM_ZU}{X^TM_ZX}$ where $M_Z = I - Z^T(Z^TZ)^{-1}Z^T$ (residual making matrix). Noting that $Cov(X,Z) = c_3+c_1c_I$ and $Cov(U,Z) = c_I$, we have:

\begin{align}
    \frac{X^TM_ZU}{X^TM_ZX} &= \frac{\sum_{i=1}^{n}[x_i-(c_3+c_1c_I)z_i][u_i-c_Iz_i]}{\sum_{i=1}^{n}[x_i - (c_3+c_1c_I)z_i]^2}
    &\overset{p}{\to} \frac{Cov[X-(c_3+c_1c_I)Z,U-c_IZ]}{Var[X-(c_3+c_1c_I)Z]}\\
    &= \frac{c_1 + c_3c_I - c_I(c_3+c_1c_I)}{1-(c_3+c_1c_I)^2} = \frac{c_1(1-c_I^2)}{1-(c_3+c_1c_I)^2}
\end{align}
\end{proof}

\subsubsection{Proof for the Consistency of $\hat{A}_4$}
\begin{proof}
In a similar derivation Proposition 3.1 but plugging in $c_0X + c_2U + \epsilon_3$ for the structural equation of $Y$, we obtain
\begin{align}
    \hat{A}_4 &\overset{p}{\to} c_0 + c_2\frac{Cov(Z,U)}{Cov(Z,X)} + \frac{Cov(Z,\epsilon_3)}{Cov(Z,X)}\\
    &= c_0 + c_2\frac{Cov(Z,U)}{Cov(Z,X)}\\
    &= c_0 + \frac{c_2c_I}{c_3+c_1c_I}
\end{align}
Note that the unconditional association between $X$ and $Z$ goes has two paths: the direct path $X \to Z$ and the indirect path $X \gets U \to Z$.
\end{proof}

\subsection{Proofs for Proposition \ref{rmk:em}}
\subsubsection{Proof for the consistency of $\hat{A}_2$}

\begin{proof}
We simply will compute the regression of $Y$ on $X$ where $\hat{A}_2 \overset{p}{\to} Cov(Y,X)$.
\begin{align}
    Cov(Y,X) &= Cov(\beta_1X + \beta_2U + \beta_3 XU + \epsilon_2,X) = \beta_1 + \beta_2\alpha_2 + \beta_3Cov(XU,X)\\
    &= \beta_1 + \alpha_2\beta_2 + 2\alpha_1\alpha_3\beta_3
\end{align}
Where $Cov(XU,X) = 2\alpha_1\alpha_3$ because 
\begin{align}
    Cov(XU,X) &= Cov(U(\alpha_1Z + \alpha_2U + \alpha_3ZU + \epsilon_1),X)\\
    &= \alpha_1Cov(UZ,X) + \alpha_2Cov(U^2,X) + \alpha_3Cov(ZU^2,X)\\
    &= \underbrace{\alpha_1E[UZX]}_{B_1} + \underbrace{\alpha_2E[U^2X]}_{B_2} 
    + \underbrace{\alpha_3E[ZU^2X]}_{B_3}
\end{align}
$B_1 = \alpha_1\alpha_3$ because
\begin{align}
    E[XUZ] &= E[ZE[XU|Z]] = E[ZE[(\alpha_1Z + \alpha_2U + \alpha_3ZU + \epsilon_1)U|Z]]\\
    &= E[Z(\alpha_1E[UZ|Z] + \alpha_2E[U^2|Z] + \alpha_3E[ZU^|Z] + E[U\epsilon_1|Z)]\\
    &= E[\alpha_1Z^2E[U] + \alpha_2ZE[U^2] + \alpha_3ZE[U^2] + E[U\epsilon_1]]\\
    &= E[\alpha_2Z + \alpha_3Z^2] = \alpha_3
\end{align}
$B_2 = 0$ because $E[U^2X] = E[U^2E[X|U]] = \alpha_2E[U^3] = 0$ because $E[U^3] = 0$.

$B_3 = \alpha_1\alpha_3$ because
\begin{align}
    E[ZU^2X] &= E[ZU^2(\alpha_1Z + \alpha_2U + \alpha_3ZU + \epsilon_1)]\\
    &= \alpha_1E[Z^2U^2] + \alpha_2E[ZU^3] + \alpha_3E[Z^2U^3]\\
    &= \alpha_1
\end{align}

Where the last equality follows because $Z\ind U$ and $E[U^3] = 0$.
\end{proof}

\subsubsection{Proof for the consistency of $\hat{A}_3$}

\begin{proof}
Similar to proposition 3.2, we will proceed via FWL to find the value of $A_3 = \frac{X^TM_ZY}{X^TM_ZX}$. First, we need to find $Cov(Z,X)$ and $Cov(Z,Y)$:
\begin{align}
    Cov(Z,X) &= Cov(Z,\alpha_1Z + \alpha_2U + \alpha_3ZU + \epsilon_1) = \alpha_1\\
    &\text{and}\\
    Cov(Z,Y) &= Cov(Z,\beta_1X + \beta_2U + \beta_XU + \epsilon_2) = E[Z(\beta_1X + \beta_2U + \beta_XU + \epsilon_2)]\\
    &= E[\beta_1XZ + \beta_2UZ + \beta_3XUZ + Z\epsilon_2] = \alpha_1\beta_1 + \alpha_3\beta_3
\end{align}
Going back to FWL, letting $\lambda = \alpha_1\beta_1 + \alpha_3\beta_3$, we now have
\begin{align}
    \frac{X^TM_ZY}{X^TM_ZX} &= \frac{(X-\alpha_1Z)^T(Y-\lambda Z)}{(X-\alpha_1Z)^T(X-\alpha_1Z)}
    = \frac{\sum_{i=1}^{n}X_iY_i-\lambda X_iZ_i - \alpha_1Z_iY_i + \alpha_1\lambda Z_i^2}{\sum_{i=1}^{n}(X_i-\alpha_1Z_i)^2}\\
    &\overset{p}{\to} \frac{E[X_iY_i - \lambda X_iZ_i - \alpha_1Z_iY_i + \alpha_1\lambda Z^2]}{E[(X_i-\alpha_1Z_i)^2]}\\
    &= \frac{Cov(X,Y) - \lambda Cov(X,Z) - \alpha_1 Cov(Z,Y) + \alpha_1\lambda Var(X)}{1-\alpha_1^2}\\
    &=\frac{(\beta_1 + \alpha_2\beta_2 + 2\alpha_1\alpha_3\beta_3) - \lambda\alpha_1 - \alpha_1\lambda + \alpha_1\lambda}{1-\alpha_1^2}\\
    &= \frac{\beta_1 + \alpha_2\beta_2 + 2\alpha_1\alpha_3\beta_3 - \alpha_1^2\beta_1 + \alpha_3\beta_3\alpha_1}{1-\alpha_1^2} = \beta_1 + \frac{\alpha_2\beta_2 + \alpha_1\alpha_3\beta_3}{1-\alpha_1^2}
\end{align}
\end{proof}

\subsubsection{Proof for the consistency of $\hat{A}_4$}
\begin{proof}
We can compute the Wald estimator from the quantities computed in the proof for $\hat{A}_3$. In particular, $Cov(Z,Y)$ and $Cov(Z,X)$.
\begin{align}
    \hat{A}_4 &\overset{p}{\to} \frac{Cov(Z,Y)}{Cov(Z,X)} = \frac{\alpha_1\beta_1 + \alpha_3\beta_3}{\alpha_1}
    = \beta_1 + \frac{\alpha_3\beta_3}{\alpha_1}
\end{align}
\end{proof}

\subsection{Proofs for Proposition \ref{rmk:uc_larger}}
\subsubsection{Proof for the consistency of $\hat{A}_2$}
\begin{proof}
    Our goal is to find $\frac{\partial}{\partial x}E[Y|x,w]$, which, by FWL, is equivalent to the convergence in probability of $\frac{X^TM_WY}{X^TM_WX}$. Because we have already demonstrated the use of FWL, we will skip several intermediate steps. Letting $\eta = c_6 + c_5c_0$
    \begin{align}
       \frac{X^TM_WY}{X^TM_WX} &= \frac{(X-c_5W)^T(Y-\eta W}{((X-c_5W))} \overset{p}{\to} \frac{Cov(X,Y)-\eta Cov(W,Y) + c_5\eta}{1-c_5^2}\\
       &= \frac{c_0 + c_1c_2+c_2c_3c_I - \eta c_5 - \eta c_5 + c_5\eta)}{1-c_5^2} = c_0 + \frac{c_1c_2+c_2c_3c_I}{1-c_5^2}
    \end{align}
\end{proof}
\subsubsection{Proof for the consistency of $\hat{A}_3$}
\begin{proof}
    Our goal is to find $\frac{\partial}{\partial x}E[Y|x,w,z]$, which, by FWL, is equivalent to the limit of $\frac{X^TM_\bold{B}Y}{X^TM_\bold{B}X}$ where $\bold{B} = [W,Z]$ or the residuals of regressing both $W$ and $Z$. To simplify matters, we can expand $Y$
    \begin{align}
        \frac{X^TM_\bold{B}Y}{X^TM_\bold{B}X} &= \frac{X^TM_\bold{B}(c_0 + c_2U + c_6W + \epsilon_2)}{X^TM_\bold{B}X}\\
        &= c_0 + c_2\frac{X^TM_\bold{B}U}{X^TM_\bold{B}X}
    \end{align}
    Where the last line follows from the orthogonality of $\epsilon_2$ and the result of $W$ being regressed upon itself being that the residuals are $0$. So we are focused on two regressions: $E[X|w,z] = \gamma_1W + \gamma_2Z$ and $E[U|w,z] = \pi_1W + \pi_2Z$. Using FWL, we find the value of the coefficients to be
    \begin{align}
        \gamma_1 &= \frac{W^TM_ZX}{W^TM_ZW} \overset{p}{\to} c_5 - \frac{c_1c_Ic_7}{1-c_7^2},\\
        \gamma_2 &= \frac{Z^TM_WX}{Z^TM_WZ} \overset{p}{\to} c_3 + \frac{c_1c_I}{1-c_7^2},\\
        \pi_1 &= \frac{W^TM_ZU}{W^TM_ZW} \overset{p}{\to} \frac{-c_Ic_7}{1-c_7^2},\\
        \pi_2 &= \frac{Z^TM_WU}{Z^TM_WZ} \overset{p}{\to} \frac{c_I}{1-c_7^2}
    \end{align}
    Putting this altogether, we write $A_3 = c_0 + c_2\frac{Cov(X-\gamma_1W - \gamma_2Z, U - \pi_1W - \pi_2Z)}{Var(X-\gamma_1W - \gamma_2Z)}$ (note that we are now in the limit). Expanding the numerator we get
    \begin{align}
        c_2&\bigg[Cov(X,U) - \pi_1Cov(X,W) - \pi_2Cov(X,Z) - \gamma_1Cov(W,U) + \gamma_1\pi_1\\
        &+\gamma_1\pi_2Cov(W,Z) - \gamma_2Cov(Z,U)+\gamma_2\pi_1Cov(W,Z) + \gamma_2\pi_2\bigg]
    \end{align}

    Noting that $Cov(X,U) = c_1 + c_3c_I$, $Cov(X,W) = c_5 + c_3c_7$, $Cov(X,Z) = c_3+c_1c_I+c_5c_7$, $Cov(W,U) = 0$, $Cov(W,Z) = c_7$, and $Cov(Z,U) = c_I$, we can simplify the above expression to
    \begin{align}
        &c_1c_2 + \frac{c_1c_2c_I^2c_7}{1-c_7^2} - \frac{c_1c_2c_I^2c_7}{(1-c_7^2)^2} + \frac{c_1c_2c_Ic_7^3}{(1-c_7^2)^2} - \frac{c_1c_2c_I^2}{1-c_7^2}\\
        &= c_1c_2 + \frac{c_1c_2c_I^2c_7-c_1c_2c_I^2}{1-c_7^2} + \frac{c_1c_2c_I^2c_7(c_7^2-1}{(1-c_7^2)^2} = c_1c_2 + \frac{c_1c_2c_I^2c_7-c_1c_2c_I^2-c_1c_2c_I^2c_7}{1-c_7^2}\\
        &= c_1c_2 - \frac{c_1c_2c_I^2}{1-c_7^2}
    \end{align}
    For the denominator
    \begin{align}
        Var(X-\gamma_1W - \gamma_2Z) &= 1 + \gamma_1^2 + \gamma_2^2 - 2\gamma_1Cov(X,W) - 2\gamma_2Cov(X,Z)+2\gamma_1\gamma_2Cov(W,Z)\\
    \end{align}
    After combining like terms we obtain
    \begin{align}
        &1 - c_5^2 - c_3^2 + \frac{c_1^2c_I^2}{(1-c_7^2)^2} - 2c_3c_5c_7 - 2c_1c_3c_I - 2\frac{c_1^2c_I^2}{1-c_7^2} - \frac{c_1^2c_I^2c_7^2}{(1-c_7^2)^2}\\
        &= 1 - c_5^2 - c_3^2 - 2c_3c_5c_7 - 2c_1c_3c_I - \frac{c_1^2c_I^2}{1-c_7^2} + [c_3^2c_7^2 - c_3^2c_7^2]\\
        &= 1 - (c_5 + c_3c_7)^2 + c_3^2c_7^2 - c_3^2 - 2c_1c_3c_I - \frac{c_1^2c_I^2}{1-c_7^2}\\
        &= 1 - (c_5 + c_3c_7)^2 - (1-c_7^2)(c_3^2 + \frac{2c_1c_3c_I}{1-c_7^2} + \frac{c_1^2c_I^2}{1-c_7^2})\\
        &= 1 - (c_5 + c_3c_7)^2 - (1-c_7^2)(c_3+\frac{c_1c_I}{1-c_7^2})^2
    \end{align}
\end{proof}
\subsubsection{Proof for the consistency of $\hat{A}_4$}
\begin{proof}
    Via FWL, $\frac{\frac{\partial}{\partial z}E[Y|z,w]}{\frac{\partial}{\partial z}E[X|z,w]} = \frac{}{} = \frac{Z^TM_WY}{Z^TM_WX}$. Re-using quantities from the proof for $A_3$ and, additionally $Cov(Z,Y) = c_0c_3+c_2c_I+c_6c_7+c_0c_5c_7+c_0c_1c_I$ and $Cov(Y,W) = c_0c_5 + c_6 + c_0c_3c_7$ we have
    \begin{align}
        \frac{Z^TM_WY}{Z^TM_WX} &\overset{p}{\to} \frac{Cov(Z,Y)-Cov(Z,W)Cov(Y,W)}{Cov(X,Z)-Cov(Z,W)Cov(X,W)}\\
        &= \frac{c_0c_3+c_2c_I+c_6c_7+c_0c_5c_7+c_0c_1c_I-c7(c_0c_5 + c_6 + c_0c_3c_7)}{c_3+c_1c_I+c_5c_7-c_7(c_5+c_3c_7)}\\
        &= \frac{c_0(c_3+c_1c_I-c_7^2c_3)+c_2c_I}{c_3+c_1c_I-c_7^2c_3} = \frac{c_2c_I}{c_3(1-c_7^2)+c_1c_I}
    \end{align}
\end{proof}

\subsection{Proof for Proposition \ref{rmk:em_larger}}
\subsubsection{Results for $A_2$}
\begin{proof}
    
Our goal is to find the value of $\frac{\partial}{\partial x}\frac{X^TM_\bold{C}Y}{X^TM_\bold{C}X}$ where $\bold{C} = [W,XW]$. Because $Cov(W,XW) = 0$, we can write the following linear conditional expectations
\begin{align}
    E[X|W,XW] = Cov(X,W)W + \frac{Cov(X,XW)}{Var(XW)}XW = \alpha_4W + \frac{2\alpha_1\alpha_5}{\alpha_4^2 + 2\alpha_5^2 + 1}XW,
\end{align}
\begin{align}
    E[Y|W,XW] = Cov(Y,W)W + \frac{Cov(Y,XW)}{Var(XW)}XW = (\beta_1\alpha_4 + \beta_4)W + \frac{2\alpha_1\alpha_5\beta_1 + \beta_5(\alpha_4^2 + 2\alpha_5^2 + 1)}{\alpha_4^2 + 2\alpha_5^2 + 1}XW.
\end{align}
Returning back to the FWL expression, we now plug in the above conditional expectations. Focusing on the numerator, which is rewritten as $Cov(X-E[X|W,XW],Y-E[X|W,XW])$, we have
\begin{align}
    &Cov(X-\alpha_4W - \frac{2\alpha_1\alpha_5}{\alpha_4^2 + 2\alpha_5^2 + 1}XW,Y - \beta_1\alpha_4 + \beta_4W - \frac{2\alpha_1\alpha_5\beta_1 + \beta_5(\alpha_4^2 + 2\alpha_5^2 + 1)}{\alpha_4^2 + 2\alpha_5^2 + 1}XW\\
    &= Cov(X,Y) - \frac{\frac{2\alpha_1\alpha_5\beta_1 + \beta_5(\alpha_4^2 + 2\alpha_5^2 + 1)}{\alpha_4^2 + 2\alpha_5^2 + 1}Cov(X,XW)}{\alpha_4^2 + 2\alpha_5^2 + 1} - \alpha_4Cov(W,Y) -\\ &\frac{2\alpha_4\alpha_5Cov(XW,Y)}{\alpha_4^2 + 2\alpha_5^2 + 1} + \frac{\frac{2\alpha_1\alpha_5\beta_1 + \beta_5(\alpha_4^2 + 2\alpha_5^2 + 1)}{\alpha_4^2 + 2\alpha_5^2 + 1}2\alpha_1\alpha_5}{\alpha_4^2 + 2\alpha_5^2 + 1}\\
    &= \beta_1 + \beta_2\alpha_2 + 2\alpha_1\alpha_3\beta_3 + \beta_4\alpha_4 + 2\alpha_1\alpha_5\beta_5 - \alpha_4(\beta_1\alpha_4+\beta_4)-\frac{2\alpha_1\alpha_5\frac{2\alpha_1\alpha_5\beta_1 + \beta_5(\alpha_4^2 + 2\alpha_5^2 + 1)}{\alpha_4^2 + 2\alpha_5^2 + 1}}{\alpha_4^2 + 2\alpha_5^2 + 1}\\
    &= \beta_1 + \beta_2\alpha_2 + 2\alpha_1\alpha_3\beta_3 - \alpha_4^2\beta_1 - \frac{4\alpha_1^2\alpha_5^2\beta_1}{\alpha_4^2 + 2\alpha_5^2 + 1}.
\end{align}
Focusing on the denominator:
\begin{align}
    &Var(X-\alpha_4W - \frac{2\alpha_1\alpha_5}{\alpha_4^2 + 2\alpha_5^2 + 1}XW) = 1 + \alpha^4 + \frac{4\alpha_1^2\alpha_5^2}{\alpha_4^2+2\alpha_5^2+1} - 2\alpha_4^2 - \frac{8\alpha_1^2\alpha_5^2}{\alpha_4^2+2\alpha_5^2+1}\\ 
    &= 1 - \alpha_4 - \frac{4\alpha_1^2\alpha_5^2}{\alpha_4^2+2\alpha_5^2+1}.
\end{align}
Thus, we obtain the final result $\beta_1 + \frac{\beta_2\alpha_2 + 2\alpha_1\alpha_3\beta_3}{1 - \alpha_4 - \frac{4\alpha_1^2\alpha_5^2}{\alpha_4^2+2\alpha_5^2+1}}$.
\end{proof}
\subsubsection{Results for $A_4$}
Due to there being two endogenous variables, $X$ and $XW$, we will need to utilize two instruments, which are $Z$ and $ZW$, respectively. We will have the following series of linear conditional expectations, noting that $Cov(XW,W) = 0$
\begin{align}
    E[X|W,Z,ZW] = Cov(X,W)W + Cov(Z,X)Z + Cov(ZW,X)ZW,
\end{align}
\begin{align}
    E[XW|Z,ZW] = Cov(XW,Z)Z + Cov(XW,ZW)ZW,
\end{align}
\begin{align}
    E[Y|\hat{X},\hat{XW},W] = \frac{Cov(Y,\hat{X})}{Var(\hat{X})}\hat{X} + \frac{Cov(Y,\hat{XW})}{Var(\hat{XW})}\hat{XW} + Cov(Y,W)W
\end{align}
where $\hat{X}$ and $\hat{XW}$ are the fitted values from the $E[X|W,Z,ZW]$ and $E[XW|Z,ZW]$, respectively. The coefficient of interest from 2SLS is thus $\frac{Cov(Y,\hat{X})}{Var(\hat{X})}$ or, returning to FWL, denoted as $\frac{\partial}{\partial \hat{x}}\frac{\hat{X}M_{\bold{D}}Y}{\hat{X}M_{\bold{D}}\hat{X}}$ where $\bold{D} = [W,\hat{XW}]$.

First, we will write out the relevant covariances and variances:
\begin{align}
    Cov(W,\hat{X}) &= \alpha_4\\
    Cov(\hat{XW},\hat{X}) &= 2\alpha_1\alpha_5\\
    Cov(W,Y) &= \beta_1\alpha_4 + \beta_4\\
    Cov(\hat{XW},Y) &= \alpha_5\beta_1Cov(Z,X) + \alpha_5\beta_3Cov(Z,XU) + \beta_5\alpha_5Cov(Z,XW)\\ &+ \alpha_1\beta_1Cov(ZW,X) + \alpha_1\beta_3Cov(ZW,XU) + \alpha_1\beta_5Cov(ZW,XW)\\
    &= 2\alpha_1\alpha_5\beta_1 + \alpha_3\alpha_5\beta_3 + \beta_5\alpha_5^2 + \alpha_1^2\beta_5\\
    Var(\hat{X)} &= \alpha_4^2 + \alpha_1^2 + \alpha_5^2\\
    Var(\hat{XW}) &= \alpha_5^2 + \alpha_1^2\\
    Cov(\hat{X},Y) &= \alpha_4Cov(W,Y) + \alpha_1Cov(Z,Y) + \alpha_5Cov(ZW,Y)\\
    &= \alpha_4(\beta_1\alpha_4 + \beta_4) + \alpha_1(\alpha_1\beta_1+\alpha_3\beta_3 + \alpha_5\beta_5) + \alpha_5(\beta_1\alpha_5 + \beta_5\alpha_1)\\
    Cov(ZW,Y) = \beta_1\alpha_5 + \beta_5\alpha_1
\end{align}

Because $Cov(\hat{XW},W) = 0$ we can simply substitute the covariances in the FWL expression as such
\begin{align}
    \frac{Cov(\hat{X}-\alpha_4W-\frac{2\alpha_1\alpha_5}{\alpha^5 + \alpha_1^2}\hat{XW},Y-(\beta_1\alpha_4 + \beta_4)W-Cov(\hat{XW},Y)\hat{XW})}{\hat{X}-\alpha_4W-\frac{2\alpha_1\alpha_5}{\alpha^5 + \alpha_1^2}\hat{XW}}
\end{align}
Now focusing on the numerator, can further simplify to
\begin{align}
    &Cov(\hat{X},Y) - (\beta_1\alpha_4 + \beta_4)Cov(\hat{X}W) - \frac{Cov(\hat{XW},Y)}{\alpha_5^2 + \alpha_1^2} - \frac{2\alpha_1\alpha_5Cov(\hat{XW},Y)}{\alpha^5 + \alpha_1^2}\\ 
    &+ \frac{2\alpha_1\alpha_5Cov(\hat{XW},Y)Var(\hat{XW})}{(\alpha^5 + \alpha_1^2)^2}\\
    &= \alpha_1(\alpha_1\beta_1+\alpha_3\beta_3 + \alpha_5\beta_5) + \alpha_5(\beta_1\alpha_5 + \beta_5\alpha_1) - \frac{2\alpha_1\alpha_5Cov(\hat{XW},Y)}{\alpha^5 + \alpha_1^2}\\
    &= \alpha_1\beta_1 + \alpha_1\alpha_3\beta_3 + 2\alpha_1\alpha_5\beta_5 + \beta_1\alpha_5^2 \\
    &- \frac{4\alpha_1\alpha_5^2\beta_1 + 2\alpha_1\alpha_3\alpha_5^2\beta_3 + 2\alpha_1\alpha_5^3\beta_5+2\alpha_1^3\alpha_5\beta_5}{\alpha_5^2 + \alpha_1^2}\\
    &= \beta_1\bigg[\alpha_1^2 + \alpha_5^2 - \frac{4\alpha_1^2\alpha_5\beta_1}{\alpha^5 + \alpha_1^2}\bigg] + \alpha_1\alpha_3\beta_3 - \frac{2\alpha_1\alpha_3\alpha_5^2\beta_3}{\alpha_1^2 + \alpha_5^2}.
\end{align}

For the denominator, we have
\begin{align}
    Var(\hat{X}-\alpha_4W - \frac{2\alpha_1^2\alpha_5^2}{\alpha_1^2 + \alpha_5^2}\hat{XW}) &= \alpha_4^2 + \alpha_1^2 + \alpha_5^2 + \alpha_4^2 + \frac{4\alpha_1^2\alpha_5^2}{\alpha_1^2 + \alpha_5^2}\\ 
    &- 2\alpha_4^2 - \frac{8\alpha_1^2\alpha_5^2}{\alpha_1^2 + \alpha_5^2}\\
    &= \alpha_1^2 + \alpha_5^2 - \frac{4\alpha_1^2\alpha_5^2}{\alpha_1^2 + \alpha_5^2}.
\end{align}

Therefore, all together we obtain the final result of
\begin{align}
    \beta_1 + \frac{\alpha_1\alpha_3\beta_3 - \frac{2\alpha_1\alpha_3\alpha_5^2\beta_3}{\alpha_1^2 + \alpha_5^2}}{\alpha_1^2 + \alpha_5^2 - \frac{4\alpha_1^2\alpha_5^2}{(\alpha_1^2 + \alpha_5^2})}
\end{align}

\subsection{Derivation for Reduced Form Coefficient in Exclusion Restriction $\beta^R_{ER}$}
\begin{proof}
    We will first derive the proof in the case with no covariates in the DAG (i.e. $E[Y|X,Z|$) as it is more tractable to show and describe how it applies when $W$ is present. The coefficient takes the form $\frac{\operatorname{Cov}(Y - E[Y|X], Z - E[Z|X])}{\operatorname{Var}(Z - E[Z|X])}$. As shown in the previous proof, the denominator is $1-c^3$ so we will focus on the numerator.
    \begin{align*}
        \operatorname{Cov}(Y - E[Y|X], Z - E[Z|X]){\operatorname{Var}(Z - E[Z|X])} 
        &= \operatorname{Cov}(Y, Z) - \operatorname{Cov}(Y, \operatorname{Cov}(Z, X)X) \\
        &\quad - \operatorname{Cov}(\operatorname{Cov}(Y, X)X, Z) + \operatorname{Cov}(Y, X) \cdot \operatorname{Cov}(Z, X) \\
        &= c_{ER} + c_3c_0 - c_3(c_0 + c_1c_2 + c_3c_{ER}) \\
        &\quad - c_3(c_0 + c_1c_2 + c_3c_{ER}) + c_3(c_0 + c_1c_2 + c_3c_{ER}) \\
        &= c_{ER} + c_3c_0 - c_3c_0 - c_1c_2c_3 - c_3^2c_{ER} \\
        &= c_{ER}(1-c_3^2) - c_1c_2c_3
    \end{align*}
    Thus, the quantity is equal to $c_{ER} - \frac{c_1c_2c_3}{1-c^3}$. Adding $W$ to the DAG, because we are conditioning for it, there is no impact besides a portion of variation in $X$ being explained away, leading us the final quantity of $c_{ER} - \frac{c_1c_2c_3}{1-c_3^2-c_5^2}$
\end{proof}

\subsection{Proof for Independence $\phi$ Result}
\begin{proof}
    We will first translate the edge weights or the linear combination of the edge weights as $R^2$ quantities. From the main text, the initial quantity is
    \begin{align}
        \phi = \Big| \frac{c_2c_I}{c_3+\frac{c_1c_I}{1-c_7^2}}\Big|\Big|\frac{\frac{c_1c_2(1-c_7^2-c_I^2)}{1-c_7^2}}{1-(c_5+c_3c_7)^2-(1-c_7^2)(c_3+\frac{c_1c_I}{1-c_7^2})^2}\Big|^{-1}.
    \end{align}
    We can re-write the different terms as follows \citep{cinelli_making_2020}:
    \begin{align}
        |c_1| &= \sqrt{R^2_{X \sim U | Z,W}}\frac{sd(X^{\perp Z,W})}{sd(U^{\perp Z,W})}\\
        |c_I| &= \sqrt{R^2_{Z \sim U | W}}\frac{sd(U^{\perp W})}{sd(Z^{\perp W})}\\
        |c_7| &= \sqrt{R^2_{Z \sim W}}\\
        (c_5 + c_3c_7)^2 &= R^2_{X \sim W}\\
        (c_3 + \frac{c_1c_I}{1-c_7^2})^2 &= R^2_{X \sim Z|W}\frac{Var(X^{\perp W})}{Var(Z^{\perp W})}\\
        \frac{1-c_I^2-c_7^2}{1-c_7^2} &= \frac{1-R^2_{Z\sim W + U}}{1-R^2_{Z \sim W}} = 1 - R^2_{Z \sim U | W}.
    \end{align}
    Canceling out $c_2$ on both sides and substituting these quantities, we obtain
    \begin{align}
        \frac{\sqrt{R^2_{Z \sim U | W}}\frac{sd(U^{\perp W})}{sd(Z^{\perp W})}}{R^2_{X \sim Z|W}\frac{Var(X^{\perp W})}{Var(Z^{\perp W})}} = \frac{\sqrt{R^2_{X \sim U | Z,W}}\frac{sd(X^{\perp Z,W})}{sd(U^{\perp Z,W})}}{1-R^2_{X \sim W}-(1-R^2_{X \sim W})(R^2_{X \sim Z|W}\frac{Var(X^{\perp W})}{Var(Z^{\perp W})})}.
    \end{align}
    Noting that $\frac{sd(U^{\perp W,Z})}{sd(U^{\perp W})} = \sqrt{1-R^2_{Z \sim U | W}}$, we can re-arrange terms such that
    \begin{align}
        \frac{\sqrt{R^2_{Z \sim U | W}}}{\sqrt{1-R^2_{Z \sim U | W}}} = \frac{\sqrt{R^2_{X \sim U | Z,W}}sd(Z^{\perp W})sd(X^{\perp Z,W})R^2_{X \sim Z|W}\frac{Var(X^{\perp W})}{Var(Z^{\perp W})}}{1-R^2_{X \sim W}-(1-R^2_{X \sim W})(R^2_{X \sim Z|W}\frac{Var(X^{\perp W})}{Var(Z^{\perp W})})}.
    \end{align}
    Squaring both sides obtains the final result.
\end{proof}

\subsection{Proof for Heterogeneity $\phi$ Result}
\begin{proof}

From the main text, the original quantity is
\begin{align}
    \phi = \Bigg|\frac{\alpha_1\alpha_3\beta_3 - \frac{2\alpha_1\alpha_3\alpha_5^2\beta_3}{\alpha_1^2 + \alpha_5^2}}{\alpha_1^2 + \alpha_5^2 - \frac{4\alpha_1^2\alpha_5^2}{(\alpha_1^2 + \alpha_5^2})}\Bigg|\Bigg|\frac{\alpha_2\beta_2 + 2\alpha_1\alpha_3\beta_3}{1-\alpha_4^2-\frac{4\alpha_1^2\alpha_5^2}{1 + \alpha_4^2 + 2\alpha_5^2}}\Bigg|^{-1}.
\end{align}

Noting that $1-\alpha_4^2-\frac{4\alpha_1^2\alpha_5^2}{1 + \alpha_4^2 + 2\alpha_5^2} = Var(X^{\perp W,XW})$, we may simplify as follows
\begin{align}
    \frac{\frac{\beta_2 \alpha_2}{2 \alpha_1 \alpha_3 \beta_3} + 1}{Var(X^{\perp W,XW})}
    &\left(\frac{\frac{1}{2} - \frac{2 a_5^2}{a_1^2 + a_5^2}}{a_1^2 + a_5^2 - \frac{4 a_1^2 a_5^2}{a_1^2 + a_5^2}}\right)^{-1} \\
    \frac{\frac{\beta_2 \alpha_2}{2 \alpha_1 \alpha_3 \beta_3} + 1}{Var(X^{\perp W,XW})} &\left(\frac{\frac{1}{2}(a_1^2 + a_5^2) - 2a_5^2}{(a_1^2 + a_5^2)^2 - 4 a_1^2 a_5^2}\right)^{-1}\\
    \frac{\frac{\beta_2 \alpha_2}{2 \alpha_1 \alpha_3 \beta_3} + 1}{Var(X^{\perp W,XW})} &\left(\frac{\frac{1}{2}a_1^2 - \frac{3}{2}a_5^2}{(a_1^2 - a_5^2)^2}\right)^{-1}\\
    \left|\frac{\beta_2 \alpha_2}{2 \alpha_1 \alpha_3 \beta_3}\right| &\left|\frac{(\frac{1}{2}a_1^2 - \frac{3}{2}a_5^2)Var(X^{\perp W,XW}) - (a_1^2 - a_5^2)^2}{(a_1^2 - a_5^2)^2}\right|^{-1}\\
    \left|\frac{2 \alpha_1 \alpha_3 \beta_3}{\beta_2 \alpha_2}\right| & \left|\frac{(a_1^2 - a_5^2)^2}{(\frac{1}{2}a_1^2 - \frac{3}{2}a_5^2)Var(X^{\perp W,XW}) - (a_1^2 - a_5^2)^2}\right|^{-1}\\
    |\alpha_3| &\left(\left|\frac{(a_1^2 - a_5^2)^2}{(\frac{1}{2}a_1^2 - \frac{3}{2}a_5^2)Var(X^{\perp W,XW}) - (a_1^2 - a_5^2)^2}\right|\left|\frac{\beta_2 \alpha_2}{2\alpha_1 \beta_3}\right|\right)^{-1} \\
 \end{align}  

This gives us the final result in the text.
\end{proof}

\subsection{Notes about Simulations}

Specifically, we have restricted the variance for all variables to 1. Therefore, choosing values of the structural coefficients for simulations to verify our derivations is subject to constraints. Importantly, the variance of the stochastic error (e.g. $\sigma_{\epsilon_1}$ in Eq \ref{eq:simple_first}) must be chosen carefully such that we simulate the random variables properly. We detail these constraints for each scenario below, which are calculated by taking variances of $X$ and $Y$ and computing the relevant covariances.

\subsubsection{Perfect IV}
We are subject to the following constraints:
\begin{enumerate}
    \item $\sigma_X^2 = 1 = c_3^2 + c_1^2 + \sigma_{\epsilon_1}^2$
    \item $\sigma_Y^2 = 1 = c_0^2 + c_2^2 + 2c_0c_1c_2 + \sigma_{\epsilon_2}^2$
\end{enumerate}

\subsubsection{Exclusion Restriction}
Without Covariates
\begin{enumerate}
    \item $\sigma_X^2 = 1 = c_3^2 + c_1^2 + \sigma_{\epsilon_1}^2$
    \item $\sigma_Y^2 = 1 = c_0^2 + c_2^2 + c_{ER}^2 + 2c_0c_1c_2 + 2c_0c_3c_{ER} + \sigma_{\epsilon_2}^2$
\end{enumerate}
With Covariates
\begin{enumerate}
    \item $\sigma_X^2 = 1 = c_1^2 + c_3^2 + c_5^2 + \sigma_{\epsilon_3}^2$ 
    \item $\sigma_Y^2 = 1 = c_0^2 + c_2^2 + c_{ER}^2 + c_6^2 + 2c_0c_1c_2 + 2c_0c_3c_{ER} + 2c_0c_5c_6 + \sigma_{\epsilon_4}^2$
\end{enumerate}

\subsubsection{$U$ is a cause of $Z$}
Without Covariates
\begin{enumerate}
    \item $\sigma_X^2 = 1 = c_1^2 + c_3^2 + 2c_1c_3c_I + \sigma_{\epsilon_1}^2$
    \item $\sigma_Y^2 = 1 = c_0^2 + c_2^2 + 2c_0c_2(c_1+c_3c_I) + \sigma_{\epsilon_2}^2$
    \item $\sigma_Z^2 = 1 = c_I^2 + \sigma_{\epsilon_3}^2$
\end{enumerate}

With Covariates
\begin{enumerate}
    \item $\sigma_X^2 = 1 = c_1^2 + c_3^2 + 2c_1c_3c_I + 2c_3c_5c_7 + \sigma_{\epsilon_4}^2$
    \item $\sigma_Y^2 = 1 = c_0^2 + c_2^2 + 2c_0c_2(c_1+c_3c_I) + 2c_0c_6(c_5+c_3c_7) + \sigma_{\epsilon_5}^2$
    \item $\sigma_Z^2 = 1 = c_I^2 + c_7^2 + \sigma_{\epsilon_6}^2$
\end{enumerate}

\subsubsection{Treatment Effect Heterogeneity}
Without Covariates
\begin{enumerate}
    \item $\sigma_X^2 = 1 = \alpha_1^2 + \alpha_2^2 + \alpha_3^2 + \sigma_{\epsilon_1}^2$
    \item $\sigma_Y^2 = 1 = \beta_1^2 + \beta_2^2 + \beta_3^2 + 2\beta_1\beta_2\alpha_2 + 4\beta_1\beta_2\beta_3\alpha_1\alpha_3 + \sigma_{\epsilon_2}^2$
\end{enumerate}

With Covariates
\begin{enumerate}
    \item $\sigma_X^2 = 1 = \alpha_1^2 + \alpha_2^2 + \alpha_3^2 + \alpha_4^2 + \alpha_5^2 + \sigma_{\epsilon_3}^2$
    \item $\sigma_Y^2 = 1 = \beta_1^2 + \beta_2^2 + \beta_3^2 + 2\beta_1\beta_2\alpha_2 + 4\beta_1\beta_2\beta_3\alpha_1\alpha_3 + 2\beta_1\alpha_4\beta_4+4(\beta_1\beta_5\alpha_1\alpha_5)+2\beta_3\beta_5(\alpha_2\alpha_4+2\alpha_3\alpha_5)+\sigma_{\epsilon_4}^2$
\end{enumerate}

\end{document}